\documentclass[final]{clv2025}

\jvol{vv}
\jnum{nn}
\jyear{2025}

\dochead{Long Paper}

\pageonefooter{Action editor: \{action editor name\}. Submission received: DD Month YYYY; revised version received: DD Month YYYY; accepted for publication: DD Month YYYY.}

\usepackage{amsmath}
\usepackage{amssymb}
\usepackage{booktabs}
\usepackage[utf8]{inputenc}
\usepackage[T2A,T1]{fontenc}
\usepackage{float}

\runningtitle{Thermodynamics of words} 
\runningauthor{Frahm and Shepelyansky}

\def\eps{\varepsilon}

\begin{document}

\title{Thermodynamics of words} 

\author{Klaus M. Frahm\thanks{Equal contribution}$^{,1}$, 
Dima L. Shepelyansky\thanks{Corresponding author}$^{,*,1}$}

\affilblock{
    \affil{Univ Toulouse, CNRS, Laboratoire de Physique Th\'eorique,  
Toulouse, France\\\quad \email{dima@irsamc.ups-tlse.fr}}
}

\maketitle

\begin{abstract}
We introduce the word thermalization hypothesis according to which
in a literary work there are effective interactions between words
that drive them to thermalization where their word 
frequency distribution is the 
same as for the phenomenon of Rayleigh-Jeans (RJ) thermalization 
and condensation. 
In this concept, the different words play a role of classical oscillators (particles)
and their word frequencies, or related tokens, are analogous
to certain filling factors over energy state levels.
The thermalization process is characterized by the conservation of
total system energy, or total number of words/tokens,
and total probability norm, or total number of different words.
We apply this approach to analyze 16 literary works,
including Bible, complete works of Shakespeare, and 
data from  American National Corpus.
We show that for these cases the effective system energy is very low  
and thus there is the RJ condensation phenomenon with a big fraction 
of total probability concentrated at the lowest energy states 
corresponding to a big fraction of rare words appearing only once or 
twice in the whole text. 
Such a RJ condensation is observed for light propagation 
in multimode optical fibers and other physical systems. 
We relate the thermalization of words to 
the recently proposed wealth thermalization for households in world countries.
The economy concepts of Lorenz and Pareto curves, and also Zipf curves, are
applied to the word frequency distributions of the selected 16 cases
and it is shown that the thermalization theory gives a strikingly good 
description of real data. 
\end{abstract}

\section{Introduction}

The statistical properties of the distribution of word frequencies 
in literary works are actively studied by the scientific community
and are reported in various review articles and books
(see e.g. \cite{baayen,popescu,piantadosi} and Refs. therein).
A significant part of research is devoted to the analysis of
properties and origins of the Zipf frequency law in
natural language \cite{zipf1936,zipf1949}.
According to this law the word frequencies
decay algebraically with an exponent
\footnote{In this work, we consider in certain cases 
power law behaviors $\quad\sim(\ldots)^{-\alpha}\quad$ 
with $\alpha\approx 1$ and 
where the decay exponent is actually ``$-\alpha$''. However, 
for simplicity, we will call ``$\alpha$'' the {\em exponent} 
for a {\em decaying power law} even though it is the negative 
exponent.}
$\alpha \approx 1$ 
at high rank values of words ordered by frequency.
Detailed investigations of the Zipf law
are described e.g. in \cite{newman,borisnew,corral,linders} 
but still its deep origins remain
under debate \cite{piantadosi}.

To understand the properties of the whole
word frequency distribution
in literary works, we introduce
the WOrd Thermalization Hypothesis (WOTH)
according to which the semantic laws, meanings and
literary work concepts create links and interactions
between words that drive the word frequency
distribution to thermalization of
words. These links and interactions
between words describe the classical objects
and thus the physical laws of classical mechanics \cite{landau,zakharovbook}
conduct the whole system to a
stationary  thermal word frequency distribution.
According to the laws of statistical physics this
distribution is given by the  Rayleigh-Jeans (RJ) thermalization and
condensation phenomenon well studied in various physical systems.
In physics the RJ distribution describes  usually classical
particles or objects where the thermalization appears due to
nonlinear interactions between particles or due to
a system contact with an external thermal bath
(see e.g. \cite{rmtprl,ourfiber}).
In this context, it is also assumed that the system
evolution conserves two quantities (integrals of motion)
being the system energy and the probability
norm (or number of particles).

In physical systems, RJ thermalization and condensation
were found in numerical simulations and experiments
with light propagation in multimode
optical fibers \cite{picozzi1,wabnitz,picozzi2,ourfiber}.
The surprising feature of the RJ thermal distribution
is that at low total system energy
a large probability fraction (or number of particles)
is concentrated at low energy states
and only a very small fraction is located at high energy states.
In fact the RJ condensation belongs to
a more generic phenomenon of statistical mechanics known as the
constraint-driven condensation \cite{trizac,satya,marsili}
that describes
such diverse physical systems as jamming in traffic,
coalescence in granular media,  
gelation in networks \cite{satya}
and financial data analysis \cite{marsili}.

Recently it was shown that RJ thermalization and condensation
appear not only for the above physical systems but
also for the wealth inequality in the world, world countries,
stock exchange markets and the distribution of 
world Gross Domestic Product (GDP)
over countries \cite{jsp,complex}.
It was also shown that the nonlinear interactions
between system agents in social stratification models
on social networks also drive the system
to RJ thermalization and condensation \cite{jsp,stratif}.

The striking feature for the wealth distribution in world countries
is the strong inequality of wealth of households
as demonstrated e.g. in \cite{piketty1,piketty2,boston}.
Thus, according to \cite{piketty2} for the whole world 
50\% of the population owns only 2\% of total wealth, while the top 10\%
owns 75\% of total wealth, of which 38\% is held by the top 1\% alone.
The emergence of the phase of very poor households 
and the oligarchic phase of very rich ones
is naturally explained in the framework of the RJ thermalization approach. 
In this context, the wealth of a household is analogous to an energy level of a
physical system.

In this work, we expand the analogy between system energy levels
and wealth of households to the
word frequency in literary works.
All $N$ different words of a given literary work
are ordered by their frequency rank $K$ increasing from
rarest words to most frequent ones
as it is the case of wealth of households \cite{piketty1,piketty2,boston}.
In the modern Large Language Models (LLM) 
for Artificial Intelligence (AI)  English words
are related to tokens with on average of 4/3 tokens for a word \cite{token}.
The tokens have their price usually expressed in US dollars (USD). 
This price is typically determined by the word frequency in a literary work 
which is related to the amount of work and energy
for this word in the whole text.
Indeed, the word frequency is proportional to the effort of a text
writer when writing the text of a book. Thus the word frequency
determines a number of tokens in a text being analogous to
a state level energy in physical system
or wealth of a household. This analogy represents the main element
of the WOTH approach.

To verify the validity of the WOTH concept, 
we follow the approach used in \cite{jsp,complex}
for the wealth inequality in the world which relies on 
the concept of the Lorenz curve \cite{lorenz}, well known in the context 
of economy. This curve 
\cite{lorenz,boston} represents the dependence of cumulated 
normalized wealth $0\leq w \leq 1$ on the cumulated normalized fraction of
population or households $0 \leq h \leq 1$.
The  perfect equipartition of wealth corresponds to
the diagonal $w=h$ and the doubled area between diagonal
and the Lorenz curve $w(h)$ provides the Gini coefficient
$0 \leq G \leq 1$  \cite{gini,boston}.

The values of Gini coefficients $G$ for world countries in 2021 
are presented in \cite{wikigini} being in the range
$0.59 < G < 0.90$ while for the whole world it is 
$G=0.843$ (see \cite{complex} using data of \cite{piketty2}). 
High $G$ values  being close to unity correspond to
a high inequality in a country with a large fraction of poor population
and a tiny oligarchic population fraction that owns
a large fraction of the total wealth \cite{piketty1,piketty2}.
The details of the distribution of high revenues 
are typically better described by the Pareto curve \cite{pareto,yakovenko2} 
(see below for more details on its definition and relation to the Lorenz 
curve).
According to the above analogy rare words correspond to
poor households
and frequent words to rich households.
Using this analogy, we construct the Lorenz and Pareto curves for 
the word frequency for 10 literary works in English,
including Bible and complete works of Shakespeare,
and 5 works in French, German, Italian, Russian and Spanish.
We also analyze a quite large data set obtained from 
American National Corpus (ANC). 
For all these 16 cases, we show that the
WOTH concept based on the RJ thermalization and condensation
gives a strikingly good description of the
Lorenz, Pareto and Zipf distributions
(we show that the Zipf distribution is directly
related to the Pareto curve).

The paper is constructed as follows: the data sets of
literary works (books from certain authors), used in this work, 
are described in Section 2, 
basic properties of the RJ thermalization and condensation concept
are recalled in Section 3,
construction rules of Lorenz, Pareto and Zipf curves of 
the word frequency distributions and two related specific RJ theoretical 
models are explained in Section 4,
the results are presented in Section 5
and their global discussion is given in Section 6.
The Appendix provides additional figures and a further subsection that 
discusses the limits of the Zipf law and related power law to 
describe the word frequency distribution. 

\section{Data sets}

In our study, we consider 16 data sets which are 
15 literary works of different authors
and also the ANC data set openly available at \cite{anc} (with 
abbreviation ANC in Table~\ref{tab1}) for which 
we study the word frequency distributions. 
The texts of each literary work in English (10 cases) and 5 other 
languages (1 case per language) are obtained  from open sources on Internet
(including such web sites as
www.gutenberg.org; bdebooks.com;
archive.org; royallib.com; bdebooks.com;
oceanofpdf.com; standardebooks.org).

More specifically, the 10 considered works in English are:
\begin{itemize}
\item An Early Modern English translation of the Christian Bible commissioned
in 1604 by King James VI and I and published in 1611 \cite{bible}
and taken from the open source available at Project Gutenberg,
with abbreviation Bible;

\item The Complete Works of William Shakespeare by William Shakespeare
taken from the open source available at 
Project Gutenberg with abbreviation Shakespeare;

\item {\it I, Robot} by Isaac Asimov  with abbreviation Asimov; 

\item {\it The Complete Novels} of Jane Austen  with abbreviation Austen; 

\item {\it David Copperfield} by Charles Dickens  with abbreviation Dickens; 

\item {\it Moby Dick} by Herman Melville with abbreviation Melville; 

\item {\it 1984} by George Orwell with abbreviation Orwell; 

\item {\it The Lord of the Rings, 1-3} by J. R. R. Tolkien  with abbreviation Tolkien; 

\item {\it The Adventures of Tom Sawyer} by Mark Twain with abbreviation Twain; 

\item {\it  Mrs. Dalloway} by Virginia Woolf with abbreviation Woolf. 
\end{itemize}

The 5 works in the other languages are:
\begin{itemize}
\item {\it Don Quixote} by Miguel de Cervantes in Spanish with abbreviation Cervantes;

\item {\it La Divina Comedia} by Dante Alighieri in Italian with abbreviation Dante;

\item {\it Faust Part I} by Johann Wolfgang von Goethe in German with abbreviation Goethe;

\item {\it  Les Miserables} by Victor Hugo in French with abbreviation Hugo;
  
\item {\it Peace and War} by Lev Tolstoy in Russian with abbreviation Tolstoy.

\end{itemize}

The total number of words $N_w$ and the total number of different words $N$ 
are given in Table~\ref{tab1} for each literary work (see below for 
a more detailed discussion of the other entries of this table). For each work 
the text, in format epub or txt, is analyzed with the help
of Open AI Chart GPT
giving the word frequency distribution with
frequency of a given word $f_m$ and word index $m$
ordered by frequency (increasing $f_m$ with increasing $m=0,1,\ldots$).
Following the analogy with the wealth inequality analysis \cite{jsp,complex}, 
we attribute to each word with index $m$ a cumulated normalized 
household fraction $h(m)=m/N$ 
and a wealth value $w_m=f_m-1$ which represents the 
(reduced) frequency (we attribute a vanishing wealth $w_m=0$ to the words 
with minimal frequency value $f_m=1$). 
The data Lorenz, Pareto and Zipf curves are constructed
from these values as described below in Section 4. 
The raw data files for the word frequencies of all these 15 cases, 
extracted from the different books (or sets of books) given above,  
are available at the web site related to this work 
\cite{ourwebpage}. The ANC data set is available at \cite{anc}.

\section{RJ thermodynamics and condensation}

Rayleigh-Jeans (RJ) thermalization appears in certain 
nonlinear oscillator systems with given oscillator energies or 
frequencies $E_m$ (with $E_0<E_1<\ldots<E_{N-1}$). For example, 
in Refs. \cite{jsp,rmtprl,ourfiber,complex}, a 
nonlinear oscillator system with complex amplitudes $C_m(t)$ 
of the form $i\frac{\partial C_m(t)}{\partial t}=E_m C_m(t)+\beta 
F_m(\{C_l(t)\})$ was considered where $F_m(\{C_l(t)\})$ is some complicated 
nonlinear function of all the amplitudes $C_l(t)$ obtained from 
some special potential (see, e.g., \cite{jsp} for details) and 
$\beta$ is a parameter for the strength of the nonlinear perturbation. 
The particular form of $F_m(\{C_l(t)\})$ used in the above references 
ensured that in this nonlinear oscillator system there are two exact 
integrals of motion (i.e. conserved quantities with time $t$) 
which are system energy 
$\mathcal{H}=\sum_m E_m |C_m(t)|^2+\beta V(\{C_l(t)\})$ and 
probability norm $\eta=\sum_m|C_m(t)|^2$ . Here $V(\{C_l(t)\})$ is the potential 
associated with the nonlinear term $F_m(\{C_l(t)\})$. 
Typically, once the system is thermalized (at long times) 
the contribution of the potential 
is small, $V\sim \beta/N$, and we have $\mathcal{H}\approx 
\sum_m E_m |C_m(t)|^2$. 

In the context of statistical physics, 
this type of system corresponds to a special type of microcanonical ensemble 
with fixed energy $\mathcal{H}=E$ and 
the additional norm constraint that $\eta$ is conserved (typically 
the value $\eta=1$ is chosen). According to standard techniques of 
statistical physics \cite{landau}, 
such a system can be replaced (for $N\gg 1$) 
by a grand canonical ensemble. In this approach, the amplitudes $C_m(t)$ have 
(in equilibrium) a Gaussian distribution with the RJ average 
\begin{align}
\label{eqrhodef}
\rho_m=\langle |C_m(t)|^2\rangle =\frac{T}{E_m-\mu}
\end{align}
where the temperature $T$ and the chemical potential $\mu$ are 
determined to satisfy both constraints in average:
\begin{align}
\label{eqconstraints}
E=\langle\mathcal{H}\rangle=\sum_m E_m\rho_m\quad,\quad
1=\langle\eta\rangle=\sum_m \rho_m\ . 
\end{align}
Here $\rho_m$ represents the average probability norm of the state 
$m$ with energy $E_m$. 
These equations allow a unique physical solution of $\mu$ (and $T$) where 
$\rho_m>0$ for all modes and which can be computed by standard numerical 
techniques. Depending on $E<E_c=\sum_m E_m/N$ (or $E>E_c$), 
we have $\mu<E_0$ and $T>0$ ($\mu>E_{N-1}$ and $T<0$ respectively; 
see \cite{jsp} for details). 
In this work, we only discuss the physical case of positive 
temperature $T>0$ and, actually, in nearly all cases related to real 
word distribution data, we are in the low temperature regime, 
corresponding to $E_0<E\ll E_c$ with possible RJ condensation where a 
big part of probability is concentrated on the lowest modes $m$. 
In the next section, 
the construction of the associated Lorenz (and other related) 
curves for such an RJ system is defined.

The striking feature of the RJ distribution (\ref{eqrhodef})
at low energies $E \ll E_c$ is a condensation of a high fraction of 
total probability $\eta$ in the states in a vicinity of the 
energy ground state.
This phenomenon is for example found at the numerical simulations
and experiments with multimode optical fibers
as discussed in \cite{picozzi1,wabnitz,picozzi2,ourfiber}.

\section{Construction of Lorenz, Pareto and Zipf curves}

The concept of Lorenz curves \cite{lorenz,lcurve} is widely used in 
economy science where it shows for a given population the cumulated 
fraction $w$ of total wealth as a function of the cumulated fraction 
of households $h$ (the households or the associated population are 
assumed to be ordered with increasing wealth). 
The line $w=h$ corresponds to the case of perfect 
equipartition. It is useful to define the Gini coefficient $G\in[0,1]$, 
being the doubled area between the equipartition line and the 
Lorenz curve. Thus the Gini coefficient characterizes the global inequality of the wealth 
distribution, i.e. large of values of $G$ close 
to unity correspond to a strong inequality. 

In this work, we construct two types of Lorenz curves, first for real word frequency
distribution data and second for suitable RJ models with certain energy spectra 
(see below). For the real data of a certain literary work (or book), we consider  $N$ 
different words used in this book (or set of books, data sets) 
with word frequency values $f_m$ being the number of times each word is used in 
this book. We assume these frequency values being ordered with 
increasing values, $f_0\le f_1\le\ldots \le f_{N-1}$, and we 
attribute to each word a wealth or token value 
$w_m=f_m-1$ (reduced frequency) such that for words with minimal frequency 
$f_m=1$ we have a vanishing wealth $w_m=0$ (very ``poor'' words). 

Then the associated (real data) Lorenz curve $(h(m),w(m))$ 
for $N+1$ data points $m=0,1,\ldots,N$ is simply constructed 
as $h(m)=m/N$ and $w(m)=\sum_{k=0}^{m-1} w_m/(N \langle w_m\rangle)$ 
where $\langle w_m\rangle =\sum_{k=0}^{N-1} w_k/N$ is the average 
wealth or (reduced) word frequency with $w_m=f_m-1$. 
Here $h(m)$ represents the cumulative fraction of households (different 
words) at index $m$ and 
$w(m)$ is the cumulative normalized reduced frequency (wealth) fraction such 
that $h(0)=w(0)=0$ and $h(N)=w(N)=1$. In this construction, we assume 
that each word has the same uniform household fraction $1/N$.

For the RJ theory, we consider an ordered model spectrum $E_k$ of $N$ levels 
with $0=E_0<E_1<\ldots< E_{N-1}$ for large values of $N$ (e.g. $N=10^4$, 
see also below). Here, each level corresponds to a given agent $k$ 
with associated wealth $E_k$. We attribute to each agent a thermalized 
RJ probability according to (\ref{eqrhodef}) for a given 
energy value $E\in]E_0,E_{N-1}[$ which fixes the parameters $T$ and 
$\mu$ as described in the last section. Then, we define the RJ Lorenz curve as 
$N+1$ points $(h(m),w(m))$ (for $m=0,1,\ldots,N$)
with $h(m) =\sum^{m-1}_{k=0} \rho_k$ 
and $w(m) = \sum^{m-1}_{k=0} E_k \rho_k/E$, again satisfying 
$h(0)=w(0)=1$ and $h(N)=w(N)=1$
(in principle the number of agents can be larger than
the number of different words). 
This curve is invariant with respect to a simple energy 
rescaling $E_m\to\lambda E_m$, $E\to \lambda E$ (and similarly for $\mu$ 
and $T$). Therefore, the relevant 
parameter is the rescaled energy $\varepsilon = E/B$ 
where $B=E_{N-1}$ is the energy bandwidth of the spectrum.
In this theoretical RJ approach, it is assumed that each word 
is represented by a certain number
of thermalized agents distributed over the 
energy levels $E_k$ being in the energy band $[0,B]$. 

Concerning the specific choice of the RJ energy spectrum 
we consider two models. The first one is very simple with 
a constant density of energy states
$\nu(k) =d k/d E_k = const$ so that 
$0 \leq E_k=k/N < 1$ for $k=0,1,2,\ldots,(N-1)$. We call this model,
the RJ standard (RJS) model that was analyzed in detail 
in \cite{ourfiber,jsp}. 
For sufficiently small values of $\eps \ll 1$, we have 
RJ condensation in this model where 
a significant fraction of probability is concentrated on a 
small number of modes at minimal $k$ values. This description
provides Lorenz curves that usually qualitatively agree  
with  Lorenz curves from real data. 

However, to obtain a more exact agreement, we introduce 
the RJ extended (RJE) model with 
$E_k =(e^{ak/N}-1)/(e^a-1)$ and $a$ being 
an additional parameter besides $\eps$. The limit $a\to 0$ reproduces 
the RJS model 
with $E_k=k/N$ and for $a>0$ the density of states $\nu(E_m)$ is decreasing at 
high energies as $\nu(E_k)=dk/dE_k = N(e^a-1)/[a(1+(e^a-1)E_k]$. The global 
energy scale here is chosen such that $B=E_{N-1}\approx 1$ (such that 
$E\approx \eps$). For the wealth inequality, the decay of $\nu$ at 
high energies is rather natural since
there are fewer rich people at high wealth/energy values
(or smaller number of words with very high frequency $f_m=w_m + 1$).
Such a situation is also plausible for the word distribution in a data set 
where very high frequency words are rarer. 

In this work, we consider only results for the RJE model, since the 
real word distribution data requires typically quite high values of 
$a\approx 5$-$10$ and the RJS model (at $a=0$) only gives a 
very rough qualitative agreement with real data.

To compute the Lorenz curve from the RJ theory (\ref{eqrhodef}),
it is possible to use the above procedure for large but finite values of $N$, 
e.g. $N=10000$. However, here, we use the continuous limit $N\to\infty$ for 
which it is possible to replace the discrete sums for cumulated household 
and wealth fractions by continuous integrals and to work out closed 
analytic expressions \cite{complex} for different key quantities. 
For example the analytical RJE Lorenz curve (at $N\to\infty$) is given by 
\begin{align}
\label{eqwofh}
w^{\rm (RJE)}(h)&=\left(1-\frac{\mu}{\eps}\right)s(h)
+\frac{\mu h}{\eps}\quad,\quad D(a)=e^a-1\quad,\\
\label{eqsofh}
s(h)&=-\frac{1}{a}\ln\left\{1+\frac{\mu D(a)}
{1+\mu D(a)}
\left[\exp\left(\frac{a(1+\mu D(a))h}{D(a)(\eps-\mu)}\right)-1\right]
\right\}
\end{align}
where $\mu$ is determined by the implicit equation:
\begin{align}
\label{eqmu1}
1&=\frac{\eps-\mu}{a(D^{-1}(a)+\mu)}\ln\left(
\frac{ e^{-a}(1-\mu) }{-\mu}\right)\ .
\end{align}
Note that the quantity $s(h)$ in (\ref{eqsofh}) corresponds to the 
continuous index $s=m/N$ in the limit $N\to\infty$ which has 
a non-trivial dependence on the household fraction $h$. 

These analytic RJE expressions agree very well with the finite $N$ 
RJE curves provided $N$ is sufficiently large (identical on graphical 
precision for $N\ge 10^4$). 
We remind, that in the RJE approach with (very) large values of $N$, 
multiple agents (many values of the spectral index $k$) correspond 
to a given individual data point for real word reduced frequency 
values $w_m=f_m -1$ (one fixed value of $m$). 

To compare the RJE model with real data Lorenz curves, the parameter 
$\eps$ is fixed by the condition that the RJE and real data 
Lorenz curves have the same Gini coefficient $G$ and the parameter $a$ 
is determined by minimizing the geometrical curve distance between both 
curves (under the constraint of respecting the first condition). 
This procedure was already used in \cite{jsp,complex}. 

Another useful characteristics, often used in economy studies
(see, e.g., \cite{pareto,yakovenko2}), is the 
cumulative distribution function (CDF) $C(w_m)$, which 
represents the fraction of households (words) 
having a wealth $w_{m'}$ with $w_{m'}\ge w_m$. 
It can be directly obtained by drawing $C(w_m)=1-h(m)$ versus $w_m$ 
(for both cases of real data and RJ models). Usually, this distribution is 
described by the Pareto distribution representing an algebraic decay law 
for $w_m\ge w_c$ with some critical $w_c$ (and $C(w_m)=1$ for $w_m<w_c$). 
In our work, we simply use the notation 
{\em Pareto distribution (or curve)}  for the CDF $C(w_m)$ as such, 
also for more general cases even if there is no simple power law. 

In contrast to the Lorenz curve, the Pareto curve 
is not scale independent and in order to compare real data with 
possible RJE Pareto curves, it is necessary to fix the global wealth 
scale either by the maximal or the average value of $w_m$. 
In this work, we show 
$C(w_m)$ as a function of $w_m/\langle w_m\rangle$ where for the 
RJE model, we simply have $\langle w_m\rangle=\eps$ (since $B=1$ and 
$E_m=w_m$). For real data the average value $\langle w_m\rangle$ 
is more reliable since the maximal value of $w_m$ for a given data set 
may be subject to strong statistical fluctuations (note that $\max(w_m)=1$ 
for the RJE model since $B=1$). 

The analytical expression for the RJE Pareto curve at $N\to\infty$ is 
given by \cite{complex}:
\begin{align}
C_{\rm RJE}(w_m)&
\label{eqCw1}
=1-\frac{\eps-\mu}{a(D^{-1}(a)+\mu )}
\ln\left(\frac{1-\mu^{-1}w_m}{1+D(a)w_m}\right)\ .
\end{align}
For $w_m\gg \max(-\mu,D^{-1}(a))$, one can apply an expansion in $1/w_m$ 
which gives the quite simple expression \cite{complex}: 
\begin{align}
\label{eqCw2}
C_{\rm RJE}(w_m)&\approx
\frac{\eps-\mu}{a}\left(\frac{1}{w_m}-1\right)
=A_{\rm RJE}\left(\frac{w_m}{\langle w_m\rangle}\right)^{-1}
-A_{\rm RJE}\langle w_m\rangle
\quad,\\
\label{ARJEdef}
\mbox{with}\quad A_{\rm RJE}&=\frac{\eps-\mu}{\eps a}\approx \frac{1}{a}
\end{align}
since $\eps=\langle w_m\rangle$ (for the RJE model) 
and typically $|\mu|=-\mu\ll\eps$ 
(for all cases in this work; see also below). 
The expression (\ref{eqCw2}) is valid in the full 
interval $w_m\in[w_{\min},1]$ where $w_{\min}$ is 
some value with $w_{\min}\gg \max(-\mu,D^{-1}(a))\approx e^{-a}$ 
(since also $-\mu\ll e^{-a}$ in all cases here). 

In the context of Linguistics, one typically considers the dependence 
of the word frequencies $f_m$ on the rank $K$ with $K=1$ ($K=N$) 
for the word with maximal (minimal) value of $f_m$. For example 
in \cite{piantadosi}, power law fits of the type $f_m\sim K^{-\alpha}$ 
with some fit parameter $\alpha$ (for the negative exponent) 
are performed. In order to provide a 
presentation closer to this convention, we define the quantity 
$Z=w_m/\langle w_m\rangle$ (analogous to $f_m$ apart from rescaling) 
which we consider as a function of the rescaled rank 
$K/N$. We note that the rank index $K$ is a cumulative index
and hence it is related with the cumulative
distribution $C(w_m)$. The dependence of $Z$ on $K/N$ 
is actually just the inverse Pareto function and 
the corresponding plots in double logarithmic representation are 
very similar to the Pareto curves but with the roles of $x$- and $y$-axis 
exchanged since we can identify $C(w_m)=K/N$ (the number of words having 
$w_{m'}\ge w_m$ is just the rank $K$). We call the curves 
for the dependence of $Z$ on 
$K/N$ {\em Zipf curves} in analogy to the Zipf law that 
is a decaying power law 
with exponent $\alpha=1$, even if these curves are not exact power laws. 

For the RJE model, we can in principle extract from (\ref{eqCw1}) a 
rather complicated analytic expression for the exact RJE Zipf curve 
$Z_{\rm RJE}(K/N)$ which is just the inverse function of (\ref{eqCw1}). 
However, 
for practical purposes the approximation (\ref{eqCw2}) is sufficient 
and in Zipf curve representation it becomes 
\begin{align}
\label{Zeq1}
Z_{\rm RJE}(K/N)&\approx\frac{A_{\rm RJE}}{A_{\rm RJE}\langle w_m\rangle +(K/N)}\ .
\end{align}
This expression is valid for $Z\gg e^{-a}/\eps$ implying 
$K/N\ll \eps e^{a}/a$. 

In the next section, we present figures which compare 
real data and (precise) RJE Lorenz, Pareto and Zipf curves 
for our 16 data sets. 
Later, we also discuss these results and the 
RJE approximations (\ref{eqCw2}) and (\ref{Zeq1}) in the context 
of a possible power or Zipf law (for a certain interval of parameters).

\section{Results}

\begin{table}
\caption{Word frequency (wealth) inequality indicators for the ANC data base and 15 
book writers. The columns correspond to name, 
Gini coefficient $G$, the total number of words $N_w$, 
number of different words $N$, 
frequency share (wealth) held by the bottom 50\%,
the top 10\%, and the top 1\% of words (households) for each writer/data base. 
Values of 0\% for the bottom 50\% of words indicate that more than 50\% of 
words have a vanishing frequency/wealth value $w_m=0$ 
(corresponding to $f_m=1$ since $w_m=f_m-1$). 
For comparison also the values of $G$ and the three 
wealth shares quantities for the Lorenz curves of the World wealth of 2021 
and the company market capitalization of the London Stock Exchange (LSE) 
of 2020 are given in the last two rows (values obtained from \cite{complex}). 
}
\label{tab1}
\begin{center}
\begin{tabular}{lcrrrcc}
\toprule
\textbf{Name} & $G$ & $N_w$ & $N$ & \textbf{Bottom 50\%} & \textbf{Top 10\%} & \textbf{Top 1\%} \\
\midrule
ANC & 0.967 & 18300430 & 293867 & 0.06\% & 95.6\% & 75.9\% \\
Asimov & 0.910 & 70704 & 7323 & 0.12\% & 85.9\% & 52.6\% \\
Austen & 0.937 & 722747 & 14923 & 0.35\% & 90.6\% & 64.1\% \\
Bible & 0.944 & 789668 & 12765 & 0.37\% & 92.0\% & 67.6\% \\
Cervantes & 0.944 & 377633 & 22705 & 0.07\% & 91.5\% & 68.4\% \\
Dante & 0.931 & 97209 & 13603 & 0\% & 88.9\% & 59.7\% \\
Dickens & 0.934 & 356285 & 14159 & 0.50\% & 90.3\% & 65.0\% \\
Goethe & 0.926 & 77007 & 13734 & 0\% & 87.6\% & 58.7\% \\
Hugo & 0.940 & 67100 & 12865 & 0\% & 90.3\% & 63.4\% \\
Melville & 0.921 & 215304 & 17406 & 0.50\% & 87.6\% & 62.7\% \\
Orwell & 0.915 & 103731 & 8964 & 0.43\% & 86.7\% & 58.2\% \\
Shakespeare & 0.941 & 909658 & 26866 & 0.33\% & 91.1\% & 67.7\% \\
Tolkien & 0.930 & 518645 & 13928 & 0.50\% & 89.8\% & 61.1\% \\
Tolstoy & 0.929 & 451496 & 50869 & 0\% & 88.6\% & 62.6\% \\
Twain & 0.915 & 71221 & 7305 & 0.17\% & 86.8\% & 55.4\% \\
Woolf & 0.918 & 63958 & 7055 & 0\% & 87.4\% & 56.4\% \\
\midrule
World 2021 & 0.843 & -- & -- & 2\% & 75\% & 38\% \\
LSE 2020 & 0.909 & -- & -- & 0.7\% & 87.4\% & 44.7\% \\
\bottomrule
\end{tabular}

\end{center}
\end{table}

In Table~\ref{tab1}, we provide for the 16 data sets (ANC and 
the literary works described in Section~2)
the Gini coefficient $G$, 
the total number of words $N_w$ in the data set 
($N_w=\sum_m f_m=N+\sum_m w_m$), the number $N$ 
of different words and certain wealth/word inequality indicators 
which can be obtained from the Lorenz curve. These are the 
word frequency (wealth) share of bottom 50\% of words (households) 
and similarly the wealth share of top 10\% (top 1\%) of words (households). 
Note, that in certain cases the bottom 50\% share is simply 0\% since 
here the household fraction $h_{\min}$ of (different) words 
with vanishing wealth $w_m = f_m-1 =0$ (minimal frequency $f_m=1$) is larger 
than 0.5. For the other cases, this bottom 50\% share is 
between 0.06\% and 0.5\%. The top 10\% (1\%) have typically 85-95\% (52-75\%). 
The average Gini coefficient of these data sets is $G= 0.931 \pm 0.0139$. 
For comparison, the last two rows of Table~\ref{tab1} also provide $G$ 
and the three wealth inequality indicators for two other cases which 
are the wealth distribution of the world population for 2021 using 
data of \cite{piketty2} and the distribution of company 
market capitalization of the London Stock Exchange (LSE) of 2020 using 
data already analyzed in \cite{complex}. 

The wealth inequality indicators for the 16 word distributions 
are even more extreme than those for the world inequality 
in 2021 which is already characterized by a rather high 
Gini coefficient $G=0.843$ \cite{piketty2,jsp} and where 
50\% of the world population owns only 2\% of total wealth while 
10\% (1\%) of the population owns 75\% (38\%) of total wealth. 
However, the word distribution data is quite similar to the LSE 2020 data 
with $G=0.909$ and where 
50\% of bottom companies have only 0.7\% and 
10\% (1\%) of top companies have 87.4\% (44.7\%) of the total 
market capitalization. 
\bigskip

\begin{figure}[h]
\begin{center}
\includegraphics[width=0.9\columnwidth]{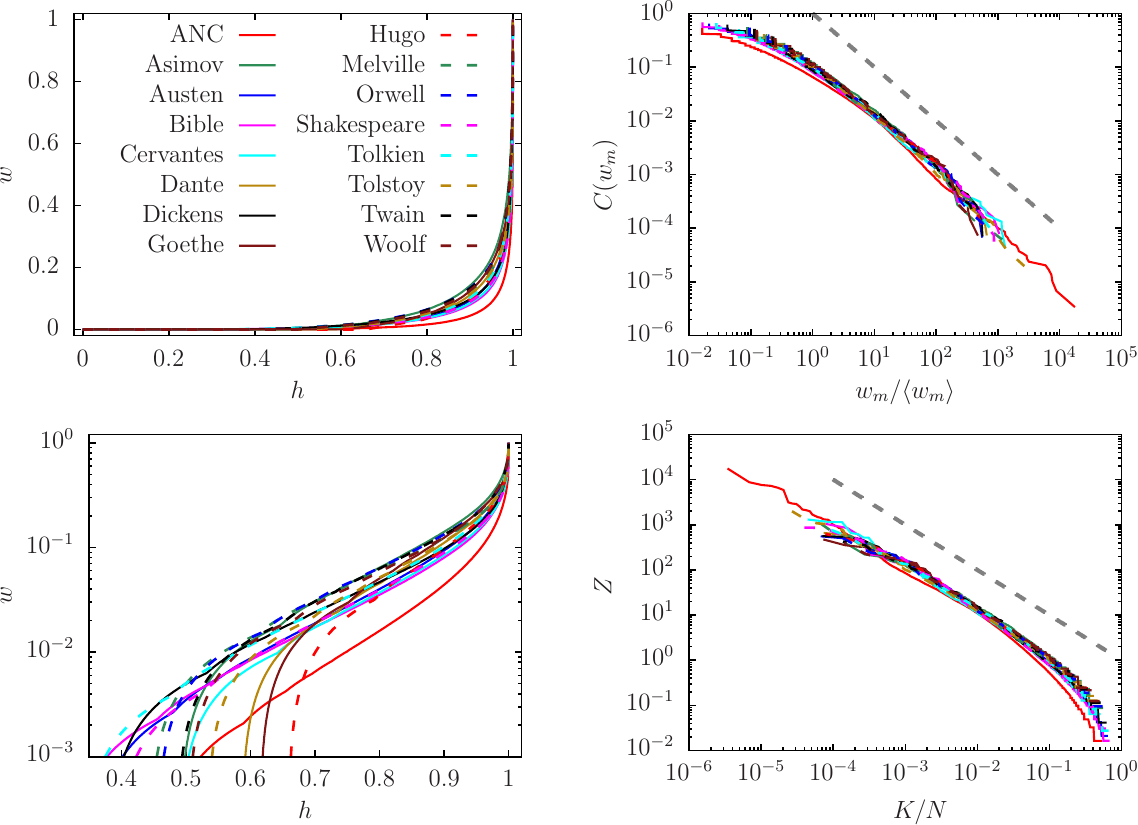}%
\end{center}
\caption{\label{fig1}
\label{fig_multi4data}
{\em Top left:} Lorenz curves for the 16 data sets of 
Table~\ref{tab1}. The $x$-axis corresponds to the 
cumulated fraction of households/words ($h$) and the $y$-axis to
the cumulated fraction of wealth/word frequency ($w$). 
The other panels show certain curves for the same data sets with 
same curve colors and types (dashed or not dashed). 
The Gini coefficients for all cases are given in Table~\ref{tab1}. 
{\em Top right:} Pareto curves $C(w_m)$ for the same data where 
$C(w_m)$ represents the fraction of words $m'$ with a word frequency 
$w_{m'}\ge w_m$ (analogous to wealth and with $w_m=f_m-1$ ). 
The $x$-axis corresponds to 
values $w_m/\langle w_m\rangle$ where $\langle w_m\rangle = N_w/N-1$ 
(see Table~\ref{tab1} for the values of $N_w$ and $N$) is the 
average reduced name frequency. Both axis are in logarithmic scale 
and the thick dashed gray line shows the power law 
$C(w_m)=(w_m/\langle w_m\rangle)^{-1}$. 
{\em Bottom left:} Lorenz curves for the same data but in logarithmic scale 
for the cumulated wealth/frequency fraction $w$ and a slightly zoomed range for 
the household variable $h\in[0.35,1]$. Note that data points with 
$h<h_{\min}$, and therefore exactly $w(h)=0$, are not visible due to the 
logarithmic scale in this panel (here 
$h_{\min}$ is the household fraction of words 
with vanishing wealth $w_m=f_m-1 = 0$, i.e. minimal frequency $f_m=1$). 
{\em Bottom right:} Zipf variable $Z=w_m/\langle w_m\rangle$ versus 
rescaled rank $K/N$ in a double logarithmic scale and where the rank 
$K=1$ ($K=N$) corresponds to the word $m$ with maximal (minimal) value of 
$w_m$. In right panels the thick dashed gray line corresponds to the power law 
$Z=(K/N)^{-1}$.
Note that $K/N$ coincides with $C(w_m)$ and 
this panel is essentially identical to the top right panel but with the 
roles of $x$- and $y$-axis exchanged.
}
\end{figure}

In Fig.~\ref{fig1} we present the Lorenz, Pareto and Zipf curves for 
all 16 data sets.
The Lorenz curve presentation emphasizes the cumulated word/wealth 
more in intermediate range of 
$h=m/N$ below unity while the Pareto and Zipf curves (both shown 
in double logarithmic scale) focus more on the very poor or very 
rich phases (minimal or maximal values of $w/\langle w_m\rangle$ 
or $K/N$). 

All 16 cases have a big phase with RJ condensation:
for 50\% of words the fraction of the whole word frequency or wealth $w$
is below 0.5\% being by a factor 4 or even more (see Table~\ref{tab1}) smaller
than for the world wealth inequality with $w=2\%$. 
The household fraction $h_{\min}$ with $w(h)=0$ (for words with 
$w_m=0$) is in the interval $0.317 \leq  h_{min} \leq 0.658$ 
for the different data set. The Lorenz curves visible 
in normal scale in the top left panel are mostly very close between 
the data sets except for the ANC Lorenz curve which is a bit below the 
other curves due to the largest Gini coefficient $G=0.967$ for this case. 
The logarithmic scale of $w(h)$ in the bottom left panel strongly enhances 
small values of the Lorenz curves, corresponding to the 
RJ condensate phase where the number of words 
is small and hence the (relative) fluctuations of Lorenz curves 
(only visible in logarithmic scale) between the 16 data sets are significant. 
However, starting from $h>0.65$, when $w > 0.01$, also the relative 
variations are reduced with rather close Lorenz curves (except for ANC) 
confirming the observation of the top left panel
demonstrating the presence of a global stationary steady-state distribution
of cumulated word frequency (or wealth). The visible deviation
from this stationary state for the ANC case can be attributed to
the fact that the ANC data is a composition 
of many literary works written by many different writers 
and containing a huge number of different words $N\approx 3\times 10^5$ 
in comparison to $N\approx 10^4$ for most other cases (the values of 
$N\approx 2\times 10^4$ for Cervantes, Shakespeare and $N\approx 5\times 10^4$ 
for Tolstoy are slightly larger but still clearly below the ANC value).
Indeed, the data of Table~\ref{tab1} shows that the Gini coefficients of 
the books of given writers are in the narrow range $0.915 \leq G\leq 0.944$
while for the ANC data set the value $G=0.967$ is visibly higher. 
This can be a result of mixing of works/books of very different authors
while each author is characterized by his/her own style of
writing. 

The global stationary steady-state distribution, with close curves 
between the different data sets, is also visible in both right panels 
of Fig.~\ref{fig1} for 
the Pareto and Zipf curves, again with slightly stronger deviations for the 
ANC curves (which also extend to a larger interval for smaller values of 
$C(w_m)=K/N$). However, to observe this universality, 
it is important that all curves are shown in the appropriate units, 
i.e. $C(w_m)$ is shown versus the rescaled wealth 
$w_m/\langle w_m\rangle $ or $Z$ versus the rescaled rank index 
$K/N$. Here the average wealth $\langle w_m\rangle=N_w/N-1$ 
takes for the 16 data sets values in the interval $[4.22,61.3]$. 
We attribute this ubiquity of word frequency distribution  
to the universal feature of RJ thermalization and condensation
based of statistical laws of physics.

In the right panels of Fig.~\ref{fig1}, 
the gray dashed (thicker) line gives for illustration the Zipf law 
$Z=N/K$ (with a certain fixed coefficient). 
The real data Pareto and Zipf curves are approximately 
parallel to this line for some limited interval, indicating 
$Z_{\rm data}\approx A(N/K)$ with $A$ typically being slightly larger 
than 0.1 according to the figure. 
Therefore, globally these curves show roughly the usual Zipf and Pareto 
algebraic dependence that captures the global probability decay of data
even if there is a certain curvature with certain deviations from the Zipf 
law which will be discussed later in more detail. 

Below, we discuss more specifically the properties of
Lorenz, Pareto and Zipf distributions
for 3 literary works from Table~\ref{tab1} 
and in particular, we compare them with the theory results of
the RJE thermalization model.
The comparison for the other 13 cases is presented in
Appendix Figs.~\ref{figA1} -- \ref{figA13}.

As explained in the previous Section, we determine 
the optimal RJE parameters $a$, $\eps$ 
for the 16 data sets. The obtained values of the parameter $a$ are rather 
high, typically $a=5$-$11$, the values of $\eps$ are quite small 
with $\eps\sim 10^{-5}$-$10^{-3}$ (specific values 
are given in the Figures below and in the Appendix).
The related values of $-\mu$ are even 
smaller (typically $-\mu/\eps\sim 10^{-3}$) indicating that all cases 
correspond to a strong RJ condensation. We note that it is possible to 
show from (\ref{eqmu1}) that $0<-\mu\ll \eps$ if $\eps\ll 1$ 
(see \cite{complex}). The two maximal values $a=10.7$ and $a=9.05$ are 
obtained for ANC or the Bible respectively. 

\begin{figure}[h]
\begin{center}
\includegraphics[width=0.9\columnwidth]{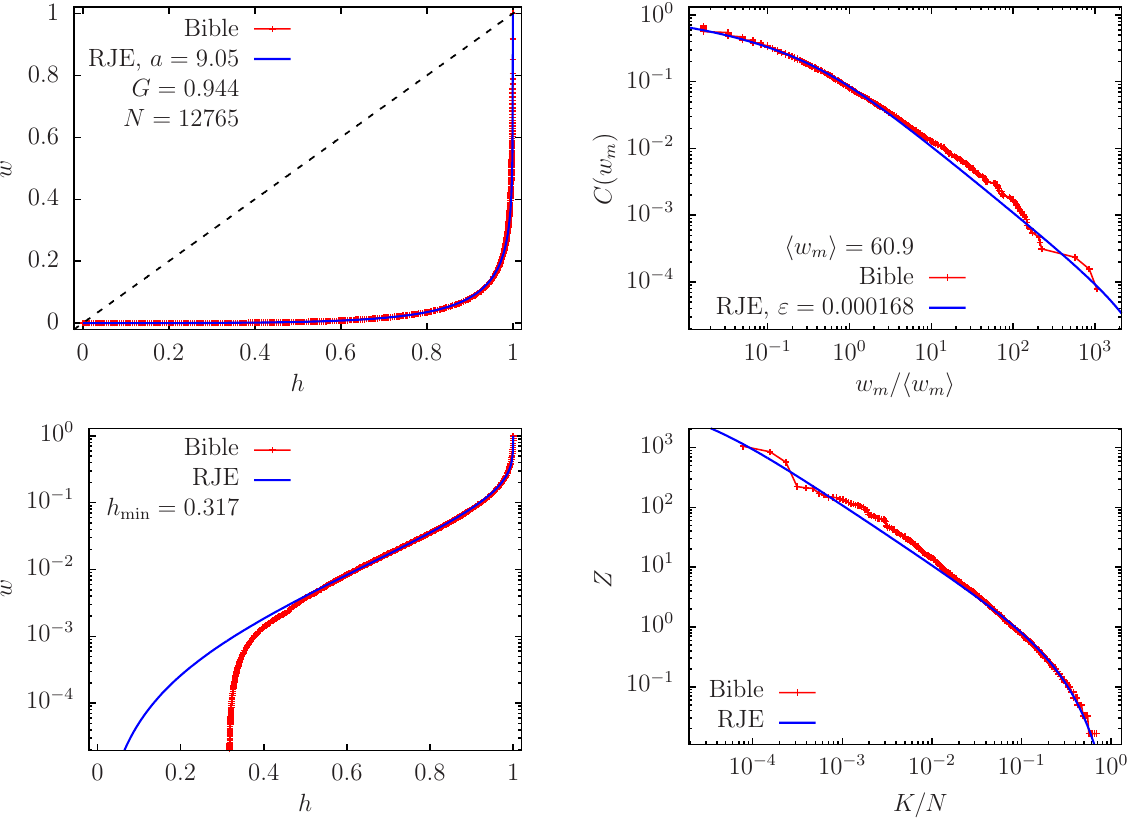}%
\end{center}
\caption{\label{fig2}
\label{fig_Bible_multi4}
All panels are in the same style as the corresponding
panels in Fig.~\ref{fig1}
but only for two curves in each panel being the
data curve for the King James Bible (red data points)
and the corresponding optimal RJE curve shown in blue.
Furthermore, in the top left panel for the Lorenz curves
also the (dashed black) line of perfect equipartition $w=h$ is shown.
The two top panels also give the values of $G$, $N$,
the real data average $\langle w_m\rangle$ and the two optimal
RJE parameters $a$ and $\eps$.
The Lorenz curves in the bottom left panel are shown in
logarithmic scale for the cumulated frequency/wealth fraction $w$ and in full
(normal) scale for the word/household variable $h\in[0,1]$.
Due to the logarithmic scale the (red)
data points for $h<h_{\min}$ with $w(h)=0$ are not visible
and the range of $w$ is chosen such that all other
data points with $w(h)>0$ are visible (the value
of $h_{\min}$ is given in the panel). }
\end{figure}

Fig.~\ref{fig2} compares real data Lorenz, Pareto and Zipf curves for 
the case of the (King James) Bible (red data points) with the 
corresponding theoretical RJE curves (blue lines) 
given by (\ref{eqwofh}) for the RJE Lorenz curve or 
(\ref{eqCw1}) for the RJE Pareto curve (the theoretical RJE Zipf curve 
is just the inverse function of (\ref{eqCw1})). The parameter values 
of $G$, $\langle w_m\rangle$ (for data), $a$, $\eps$ and $h_{\min}$ are given 
in the panels of this figure. 
The two Lorenz curves in normal representation (top left panel) for data and 
the RJE model agree nearly perfectly on graphical precision. 
In logarithmic scale (bottom left panel), there are some deviations for 
very small values $w<10^{-3}$ which are more of a technical nature due 
to the low frequency statistics of rare words. 

We note that in the RJE model the global wealth/energy scale is 
arbitrarily fixed at $\max(w_m)=\max(E_m)=B=1$ and 
gives $\langle w_m\rangle=\eps$ which is well below unity (see last section). 
Therefore, the real data and 
RJE Pareto curves have to be compared for the proper rescaled 
quantity $w_m/\langle w_m\rangle$, i.e. the wealth is measured in 
both cases (data and RJE model) in units of its average. 
Using this rescaling convention, there is also a strikingly good agreement 
between the data and the corresponding RJE Pareto and Zipf curves 
(right panels) on the whole scale of 
4 orders of magnitude for these distributions. 
We stress that the Pareto and Zipf curves have
a well visible curvature and cannot be described
by a simple algebraic decay with a certain 
exponent $\alpha \approx 1$ {\em for the full interval}. 
However, this curvature is well captured and described by the RJE theory.

\begin{figure}[h]
\begin{center}
\includegraphics[width=0.9\columnwidth]{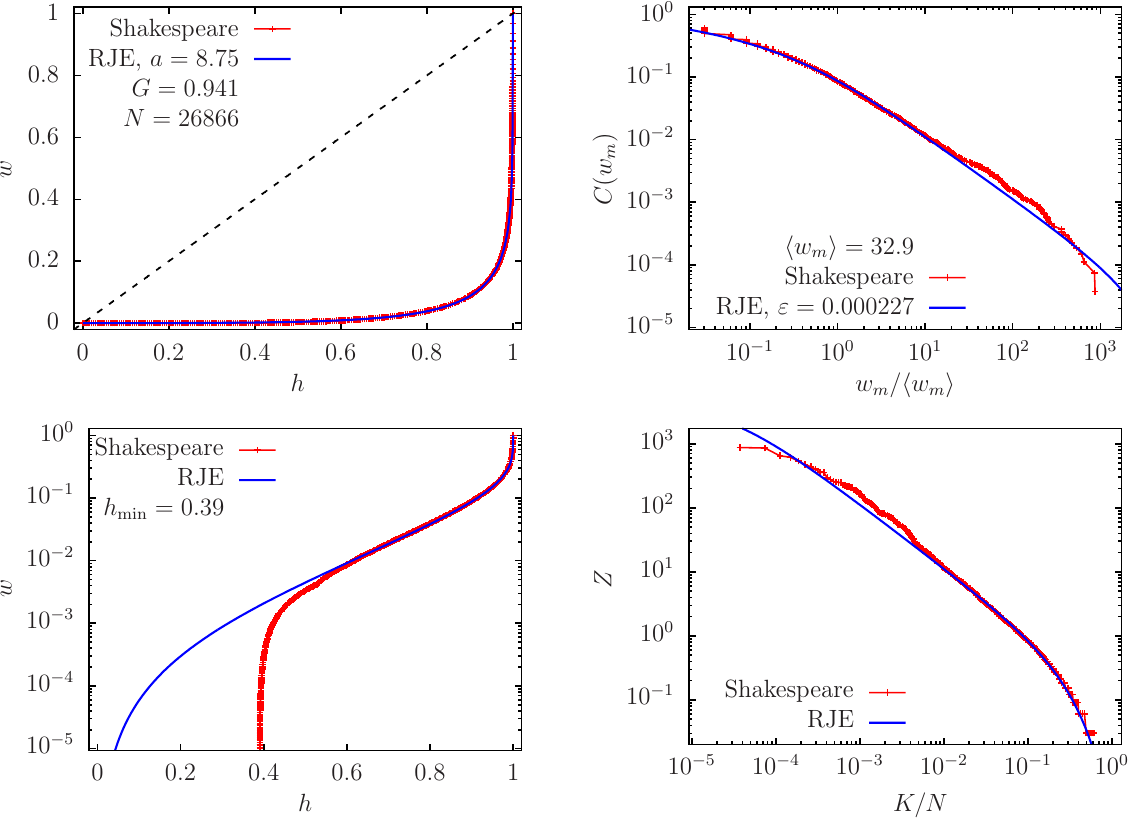}%
\end{center}
\caption{\label{fig3}
\label{fig_Shakespeare_multi4}
As Fig.~\ref{fig2} for Shakespeare.
}
\end{figure}

Fig.~\ref{fig3} is similar to Fig.~\ref{fig2} but for the complete works 
of William Shakespeare. Here, as for the case of the Bible, we also
find a very good agreement of the RJE theory with the real data. 
For the Lorenz curves there are only technical deviations (of relative 
values) for the range of very small cumulated wealth values 
$w<10^{-3}$, only visible in logarithmic scale, while 
in normal scale both curves are essentially identical. Also the Pareto 
and Zipf curves agree well between data and the RJE model in a range of 4 
orders of magnitude. Here the obtained value of $a=8.75$ is slightly 
smaller than for the case of Bible with $a=9.05$. 

\begin{figure}[h]
\begin{center}
\includegraphics[width=0.9\columnwidth]{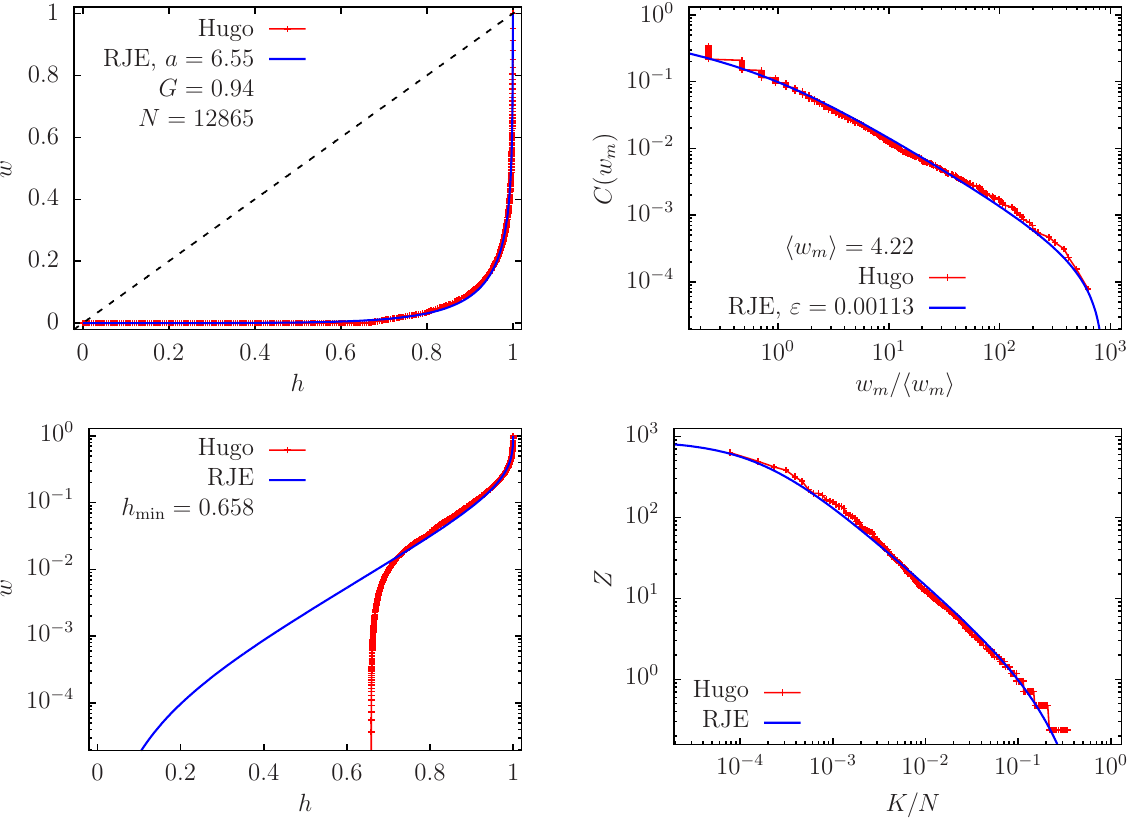}%
\end{center}
\caption{\label{fig4}
\label{fig_Hugo_multi4}
As Fig.~\ref{fig2} for Hugo.
}
\end{figure}

Fig.~\ref{fig4} compares the data of the book {\it Les Miserables} 
by Vitor Hugo with the RJE theory. 
Again, we find a very good agreement between the data and the theory.
In this case, the statistics of words $N_w$ is by a factor 10 smaller
compared to the cases of Bible and Shakespeare (see Table~\ref{tab1}).
Due to this, the small wealth deviations for the Lorenz curve 
are visible for $w<0.01$ in logarithmic scale. Here the value of 
$h_{\min}=0.658$ is very large, which affects the bottom left panel. 
However, in normal scale the data and RJE Lorenz curves are very close 
and the Pareto and Zipf curves agree on a scale of 3 orders
of magnitude. Again, the theory captures well the curved feature
of these distributions. Since the optimal value of $a=6.55$ is lower 
than in the other two cases, the curved regions of the Pareto and Zipf 
curves are relatively larger (see later discussion in the Appendix). 

In all cases of Figs.~\ref{fig2}-\ref{fig4} (and also similar 
Appendix Figs.~\ref{figA1}-\ref{figA13}), we can observe for 
the Pareto (Zipf) curve in the range of small $w_m/\langle w_m\rangle$ 
(large $K/N$) the effect, of some 
degenerate $w_m$ values being $w_m=1,2,3,\ldots$ visible as small vertical 
(horizontal) steps. 
Of course, such degeneracies mechanically reduce a bit the agreement between 
data and smooth RJE curves. But this is only a minor effect and clearly due 
to the low frequency statistics of rare words. 

We remind, that in the RJE model there are (only) two fit parameters $a$ and 
$\eps$ which are obtained by minimizing the geometrical curve distance 
between data and RJE Lorenz curves under the constraint of identical values 
of the Gini coefficient $G$. Once this fit procedure has been applied, also 
the Pareto (and Zipf) curves agree very well provided that there are shown 
in the appropriate scale (units of $\langle w_m\rangle$). Such an agreement 
is not obvious since the Pareto curve emphasizes in double logarithmic scale 
different regions for the wealth as the Lorenz curve. Furthermore, 
there are no other 
additional fit parameters for the Pareto curve, in particular the wealth units 
for the average wealth are given either directly by the data or the theory. 
We mention that the power law fits of \cite{piantadosi}, either use 
two parameters (pure power law fits) or three parameters (shifted 
power law fits).

\begin{figure}[h]
\begin{center}
\includegraphics[width=0.9\columnwidth]{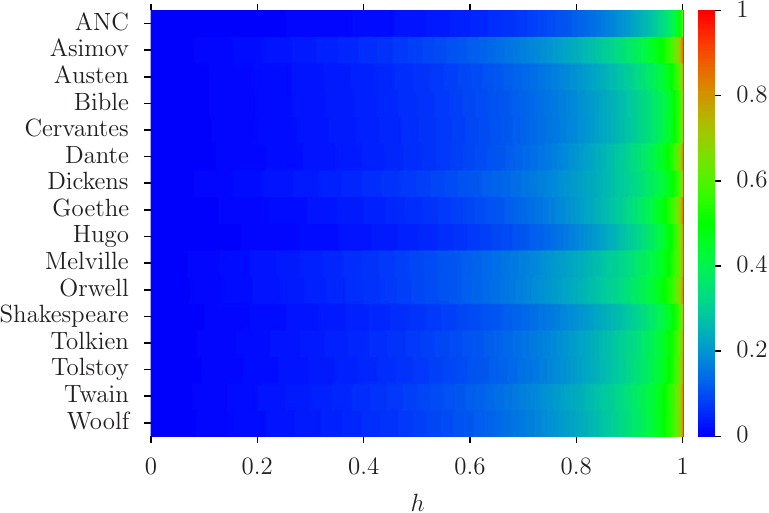}%
\end{center}
\caption{\label{fig5}
\label{fig_LOR_books}
Density color plot of Lorenz curves for the 16 cases of Table~\ref{tab1}
obtained from the RJE model with optimal RJE parameters for $\eps$ and $a$.
The $x$-axis corresponds to the 
cumulated fraction of households/words ($h$) and the $y$-axis to
data sets/writers (in alphabetical order). 
The color value corresponds to the cumulated fraction of wealth or 
rescaled word frequency ($w$). In order to enhance small 
values of $w$ a nonlinear scale 
is chosen where the values of the color bar correspond to $w^{1/2}$. 
The RJE Lorenz curves here correspond to the optimal RJE fits 
shown as blues lines in the top left panels of 
Figs.~\ref{fig2}-\ref{fig4} and
Appendix Figs.~\ref{figA1}-\ref{figA13} (see 
also these figures for the values of $\eps$ and $a$). 
}
\end{figure}

In Fig.~\ref{fig5}, we provide a global color view of 
Lorenz curves obtained within
the RJE theory for all 16 literary works of Table~\ref{tab1}.
The theory gives almost exact values of cumulated wealth $w$ 
for $w >0.01$ for all 16 cases as it is shown in
Figs.~\ref{fig2}-\ref{fig4} and Appendix Figs.~\ref{figA1}-\ref{figA13}. 
The color scale in Fig.~\ref{fig5} is nonlinear to enhance smaller wealth 
values (i.e. the numbers of the colorbar in the interval $[0,1]$ correspond 
to $w^{1/2}$ which increases the greens parts in the figure) 
but the range $w<0.01$, with possible (relative) technical deviations 
between data and RJE Lorenz curves, is still clearly in the range of 
blue color with $w^{1/2}<0.1$ for minimal values. 
Therefore, this region does not affect Fig.~\ref{fig5}.
Apart from this, there are certain variations from one case to another 
correlated with the variations of the Gini coefficient
(see Table~\ref{tab1}). For example for Asimov, Melville and Orwell, we 
have quite large green zones with (slightly) smaller values of 
$G\approx 0.91$-$0.92$ for them. On the other side, we have small 
green zones for ANC, Bible and Cervantes with (slightly) larger values 
of $G=0.967$ or $G=0.944$. 

\begin{figure}[h]
\begin{center}
\includegraphics[width=0.9\columnwidth]{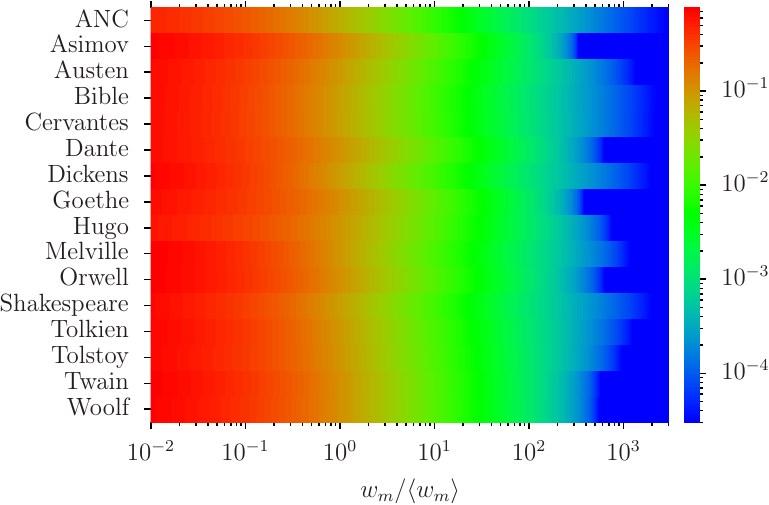}%
\end{center}
\caption{\label{fig6}
  \label{fig_Pareto_books}
Density color plot of Pareto curves for the 16 cases of Table~\ref{tab1}
obtained from the RJE model with optimal RJE parameters for $\eps$ and $a$.
The $x$-axis corresponds to the rescaled wealth $w_m/\langle w_m\rangle$ 
(in logarithmic scale) and the $y$-axis to
data sets/writers (in alphabetical order). 
The color value (in logarithmic scale) corresponds to the RJE value of 
the Pareto function $C_{\rm RJE}(w_m)$ given by Eq. (\ref{eqCw1}). 
Values of $C_{\rm RJE}(w_m)$ below the minimal color value 
$3\times 10^{-5}$, 
in particular if $C_{\rm RJE}(w_m)=0$ for $w_m>w_{\max}$, are also shown 
by blue color for the minimal value. Perfect red color corresponds 
to max$(C_{\rm RJE}(w_m))\approx 0.77$ where the maximum is taken 
over all data sets and over all $w_m$ values inside the used interval for 
$w_m/\langle w_m\rangle \in[10^{-2},3\times 10^{3}]$.
(Note for the RJE model, we have 
$w_{\max}=1$, $\langle w_m\rangle=\eps$ and 
$w_{\rm max}/\langle w_m\rangle=1/\eps$. Furthermore, the 
theoretical limit 
$C_{\rm RJE}(w_m)\approx 1$ is only reached for $w_m\ll 0.01$ not 
accessible in this figure.) 
The RJE Pareto curves here correspond to the blues lines in the top 
right panels of 
Figs.~\ref{fig2}-\ref{fig4} and \ref{figA1}-\ref{figA13} (see 
also these figures for the values of $\eps$ and $a$). 
}
\end{figure}

Fig.~\ref{fig6} gives a similar color plot view for the RJE Pareto curves 
for all 16 cases but here both $x$-axis for $w_m/\langle w_m\rangle$ 
and colorbar for $C_{\rm RJE}(w_m)$ are in logarithmic 
scale similarly as top right panels of Figs.~\ref{fig2}-\ref{fig4} and 
Appendix Figs.~\ref{figA1}-\ref{figA13} (which also show a 
very good agreement between real data and the RJE Pareto curves). 
We note that the maximum value of red color for $C(w_m)$ is 0.77 
and not unity, since this is simply 
the maximum value of $C_{\rm RJE}(w_m)$ for all 16 cases and the 
shown range $w_m/\langle w_m\rangle\ge 0.01$. The theoretical limit 
of $C_{\rm RJE}(w_m)\approx 1$ is only 
obtained for $w_m/\langle w_m\rangle\ll 0.01$, a range which is 
not visible in the figure (same effect for top right panels of other 
figures in the style of Fig.~\ref{fig2}). 
Globally, both Figs.~\ref{fig5} and \ref{fig6} illustrate quite clearly 
the universality of the RJE Pareto Lorenz and Pareto curves 
(which are all close to the corresponding real data curves). 

In this Section we presented
results for the cases of Bible, Shakespeare and Hugo
(see Figs.~\ref{fig2}-\ref{fig4}). The other 13 cases are shown in
Appendix Figs.~\ref{figA1}-\ref{figA13} which are in the same style 
way as Fig.~\ref{fig2}. 
In particular, the ANC case is presented in Fig.~\ref{figA1};
here the statistics of all words $N_w$ is very high and
the RJE theory describes well the Pareto and Zipf distributions
on a scale of 5 orders of magnitude.
The cases of Asimov, Austen, Cervantes, Dante,
Dickens, Goethe, Melville, Orwell, Tolkien, Tolstoy,
Twain and Woolf are presented in Appendix Figs.~\ref{figA2}-\ref{figA13}
respectively.
The data of all these cases agree well with the RJE theory.

We also point out that at high values of the 
parameter $a$ and low values of $\eps$ the analytic RJE expression 
for the Pareto curve gives an approximate algebraic decay of probability
with the exponent $\alpha=1$ in a certain restricted range \cite{complex} 
(see also the approximate RJE expression (\ref{eqCw2})). 
Due to the direct relation between Pareto and Zipf curves
this is also valid
for the Zipf distribution. We discuss this feature in more detail
in the Appendix with illustrations in Fig.~\ref{figA14} 
and Table~\ref{tabA1}.

\section{Discussion}

In this work, we presented the concept
of RJ thermalization and condensation
of words in literary works (or books)
of different writers and different languages.
In this WOTH approach the different $N$ words 
of a book are analogous to independent classical particles
and also analogous to households in the wealth inequality problem.
The interactions between these words (or particles)
appear due to semantic rules, meanings and writer concepts, 
generating links and nonlinear couplings between them. 
The precise form of these interactions is not specifically known but 
this is not important when applying well known concepts of statistical physics 
\cite{landau,zakharovbook}. For example, 
as for a gas of classical particles in three dimensions
almost any  weak or moderate interaction between particles
drives a system into the thermal state
with the average kinetic energy per particle 
$3kT/2$ where $k$ and $T$ are
the Boltzmann constant and gas temperature.

In the framework of the WOTH thermodynamic theory, all words in a book are 
distributed over certain states levels 
characterized by the word frequencies (and corresponding tokens).
These levels are analogous to the energy levels in a physical system
of classical particles, or to word frequencies or token levels
for a book. This is also similar to the wealth state levels of households
analyzed in \cite{jsp,complex}.
We propose the RJE model of this energy (token) spectrum
with a density of states decreasing at 
high word frequencies (or at high energies or
high wealth revenues) characterized by a certain system parameter $a$.
The thermal steady-state distribution of frequencies or tokens
over these state levels is described
by the classical RJ thermalization law (\ref{eqrhodef})
characterized by the rescaled total system energy $\eps$.

Using these RJ thermalization theory parameters $\eps$ and $a$, 
obtained by a specific fit procedure for the Lorenz curve, 
we compare the results of the RJE distribution
of word frequencies of 16 literary works
described in Section 2 and Table~\ref{tab1}.
We find that the RJE model gives a strikingly good description
of all 16 cases of real data for their Lorenz, Pareto and Zipf curves, 
and this essentially for the {\em full range} of 
the wealth or rank interval. The obtained values of the 
rescaled effective system energy $\eps$ are typically very low 
indicating that there is RJ condensation 
of a significant fraction of total probability
(fraction of all different words $N$)
at the lowest energy (token) level. This is related to 
the typically huge fraction of different words 
at lowest frequency $f_m=1$ with typical values 
of $45\%\pm 15\%$ for the considered data sets. 
Also the 50\% bottom fraction of all different words
contains only less than 0.5\% of the total
number of book words $N_w$ (see Table~\ref{tab1}).

The phenomenon of RJ condensation has also been studied previously in the 
physics of light propagation
in multimode optical fibers \cite{picozzi1,wabnitz,picozzi2}
and for the wealth inequality in the world \cite{jsp,complex}
(and also in constraint driven condensation in
other physical systems \cite{trizac,satya,marsili}).

The WOTH theory
gives a very good description
of the word frequencies in the 16 considered cases
but one can object that there are two fit parameters
in these RJE distributions. 
However, the main competing proposal to describe work frequency 
distributions based on a (shifted) power law (see e.g. \cite{piantadosi}) 
also uses two or even three fit 
parameters and even then it does not provide a good description at 
both boundaries of the interval of the frequency distribution 
(for the data used in our work). This concerns 
in particular the range of small 
wealth/word frequency or large rank index where the real data 
curves show a strong curvature for Pareto and Zipf curves 
(when shown in double logarithmic scale) 
which is very well captured by the optimal RJE curves. 
Concerning the Lorenz curves the agreement of data with 
a shifted power law proposal is even much worse. Thus in that case  the 
Lorenz curve construction as such is actually very delicate and only possible 
for the shifted power law proposal similar to (\ref{eqCw2}) 
(impossible for the simple power law). On 
the other hand the RJE Lorenz curves agree nearly perfectly 
with the data (when shown in normal scale; modest deviations for very 
small cumulated wealth values are only visible in logarithmic scale
inside strongly condensated phase). 

Apart from this, there are also important {\em conceptual features} 
that in our opinion provide a strong support to the WOTH concept:

\begin{description}
\item{F1)} the distribution of word frequencies is a
stable steady-state being very similar for all 16 cases considered
(see Figs.~\ref{fig1} and \ref{fig6}), there is also a rather
narrow interval for Gini coefficient in Table~\ref{tab1}),
this is in agreement with the thermodynamic stability of the RJ thermal
distribution;

\item{F2)} a huge condensation of rare words at
lowest word frequencies is in qualitative and quantitative
agreement with the RJ condensation phenomenon;

\item{F3)} The RJ type distributions determined by thermalization and condensation
appear in very different systems like
wealth inequality of households in the world,
stock market capitalization of companies,
GDP distribution over world countries \cite{jsp,complex}
and word frequencies in books;
the RJ condensation also describes the Lorenz curves in
various other systems (see \cite{ourwebpage});
this highlights the universal features of the RJ thermal concept.
\end{description}

We note that recently thermodynamic concepts start
to be discussed in relations to LLM
(see e.g. \cite{vetrov} and Refs. therein)
and we hope that the thermodynamics of words established here
will find useful applications there.

\appendix

\appendixsection{}

In this appendix, we present additional figures in the 
same style of Fig.~\ref{fig2} for the other data sets not shown 
in the main part and also a discussion about the limits of 
a potential power law for the Zipf curves of real data. 

\subsection{Additional figures}

\begin{figure}[H]
\begin{center}
\includegraphics[width=0.9\columnwidth]{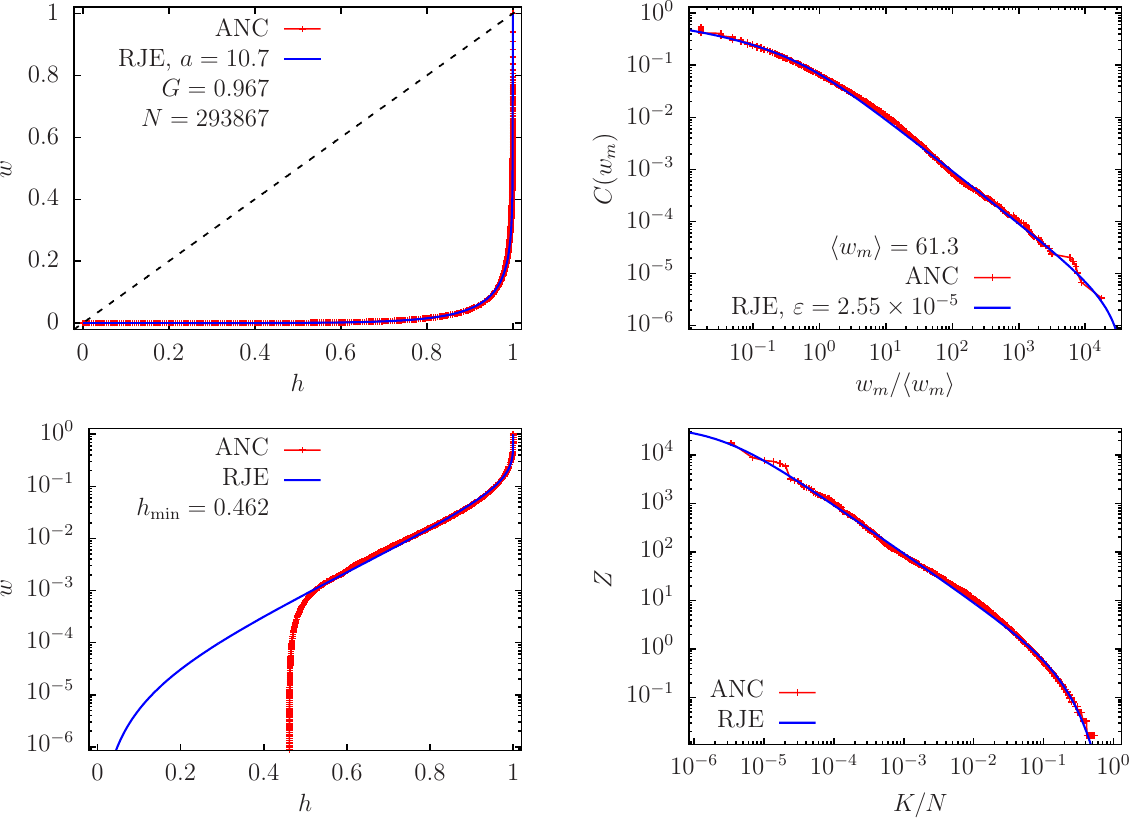}%
\end{center}
\caption{\label{figA1}
\label{fig_ANC_multi4}
As Fig.~\ref{fig2} for ANC.
}
\end{figure}

\begin{figure}[H]
\begin{center}
\includegraphics[width=0.9\columnwidth]{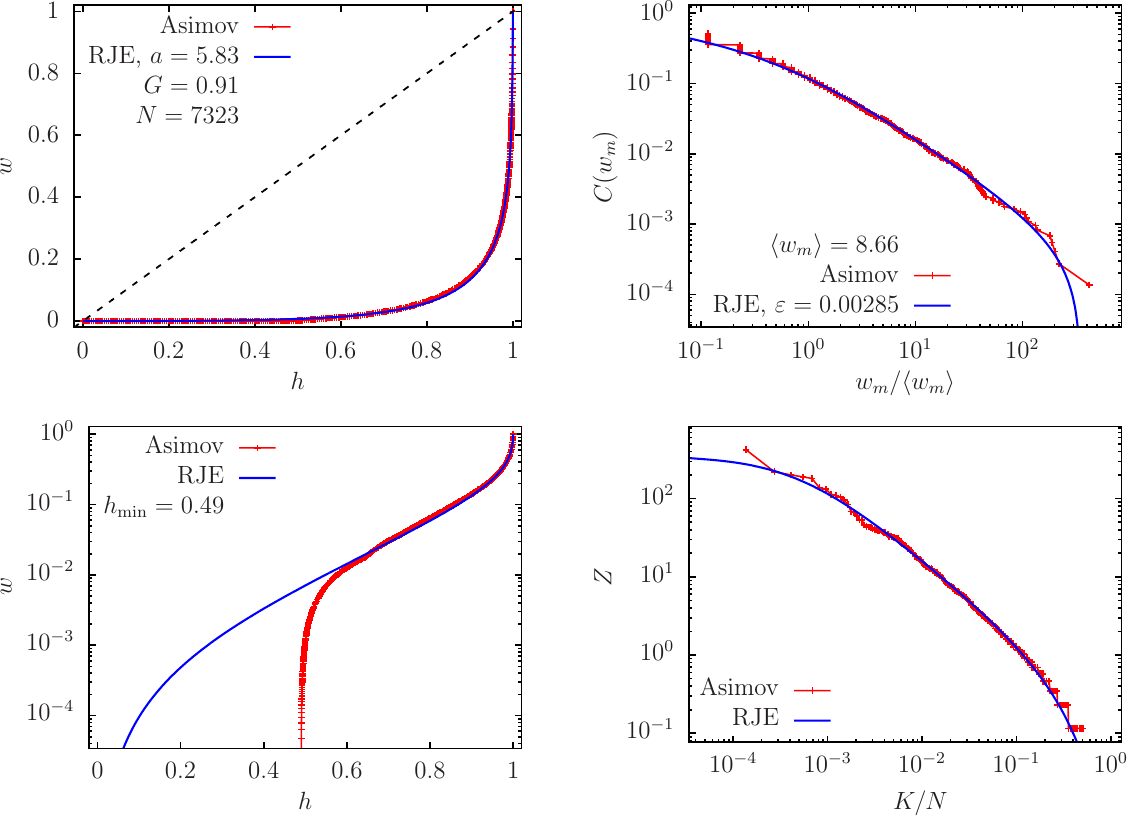}%
\end{center}
\caption{\label{figA2}
\label{fig_Asimov_multi4}
As Fig.~\ref{fig2} for Asimov.
}
\end{figure}

\begin{figure}[H]
\begin{center}
\includegraphics[width=0.9\columnwidth]{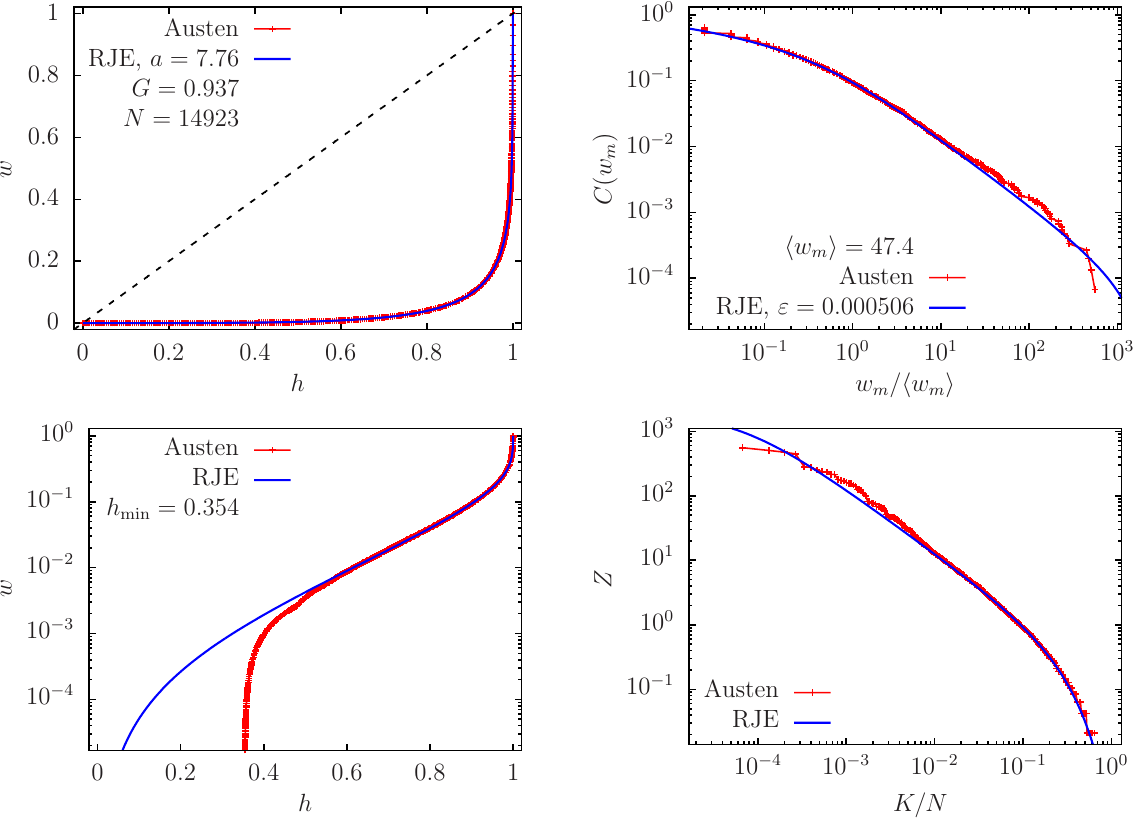}%
\end{center}
\caption{\label{figA3}
\label{fig_Austen_multi4}
As Fig.~\ref{fig2} for Austen.
}
\end{figure}

\begin{figure}[H]
\begin{center}
\includegraphics[width=0.9\columnwidth]{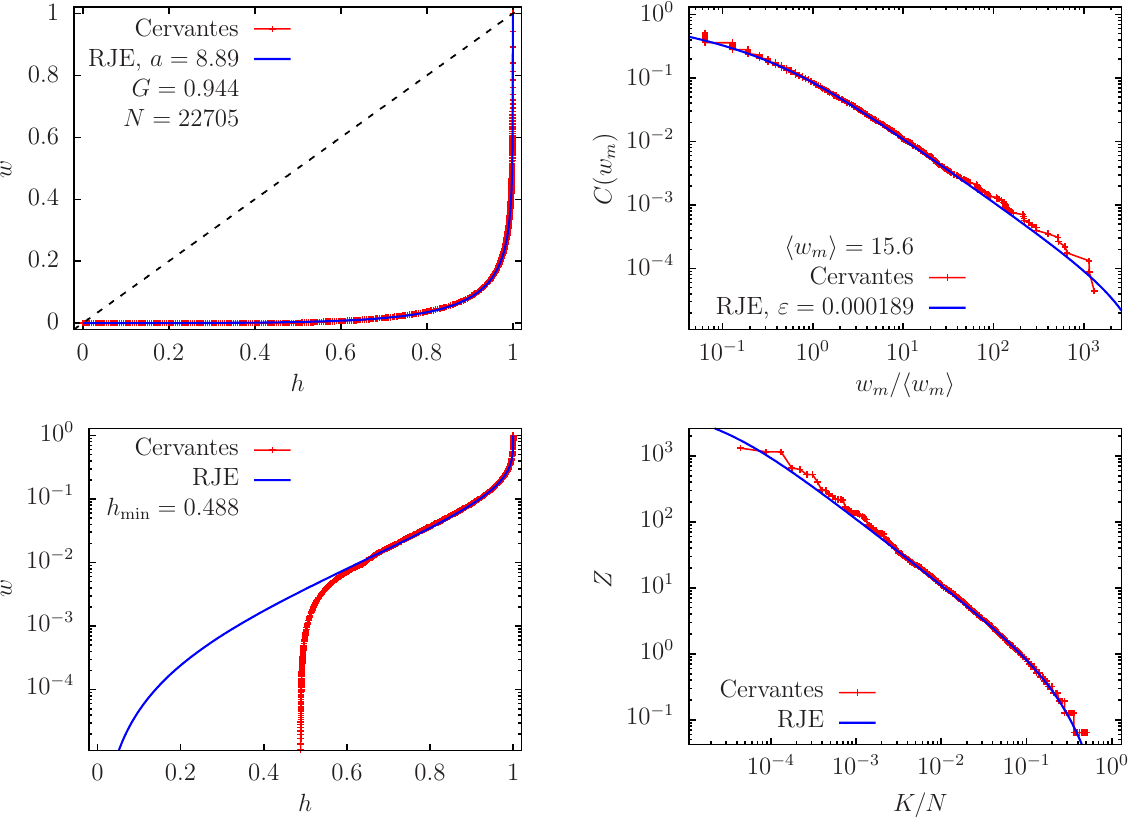}%
\end{center}
\caption{\label{figA4}
\label{fig_Cervantes_multi4}
As Fig.~\ref{fig2} for Cervantes.
}
\end{figure}

\begin{figure}[H]
\begin{center}
\includegraphics[width=0.9\columnwidth]{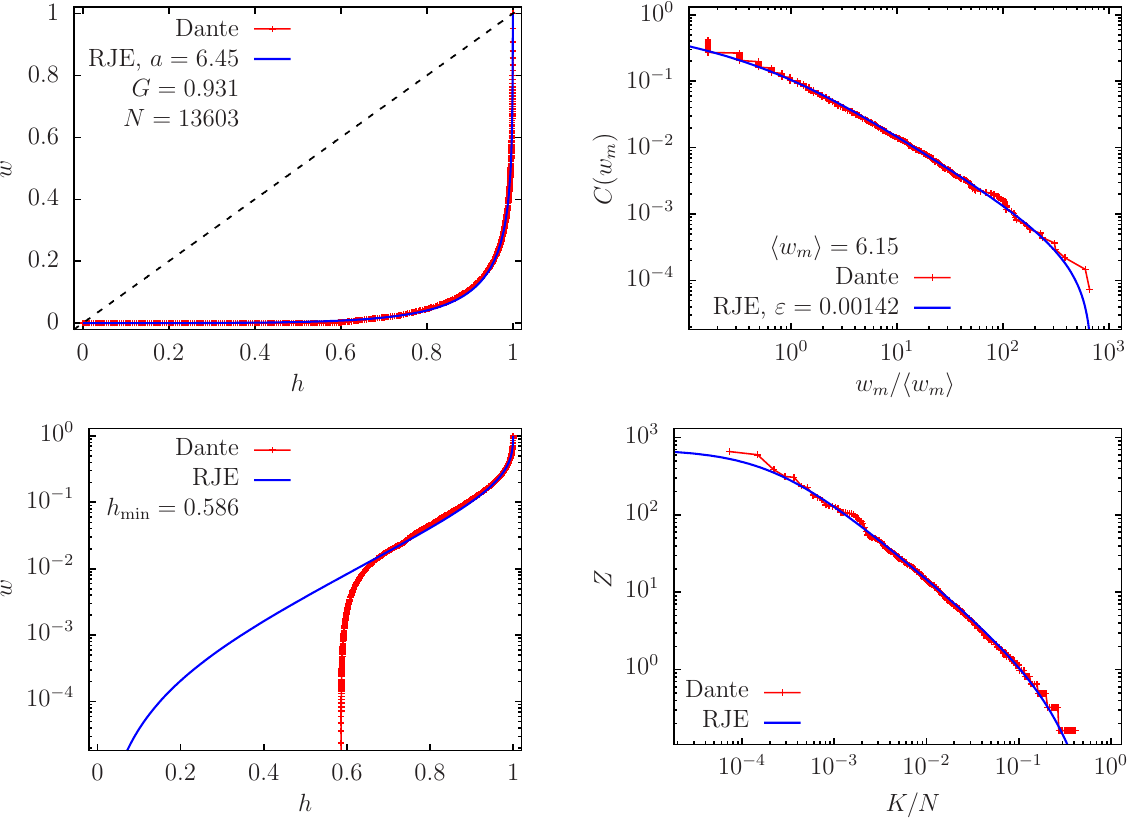}%
\end{center}
\caption{\label{figA5}
\label{fig_Dante_multi4}
As Fig.~\ref{fig2} for Dante.
}
\end{figure}

\begin{figure}[H]
\begin{center}
\includegraphics[width=0.9\columnwidth]{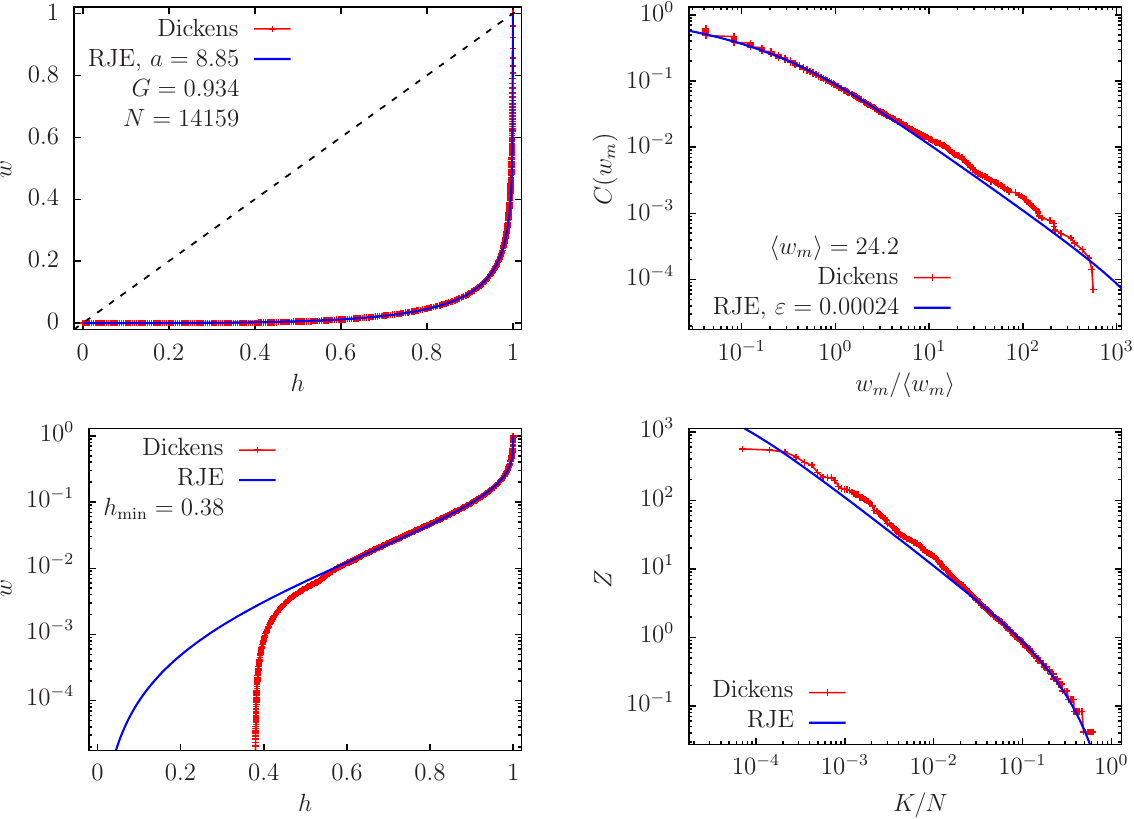}%
\end{center}
\caption{\label{figA6}
\label{fig_Dickens_multi4}
As Fig.~\ref{fig2} for Dickens.
}
\end{figure}

\begin{figure}[H]
\begin{center}
\includegraphics[width=0.9\columnwidth]{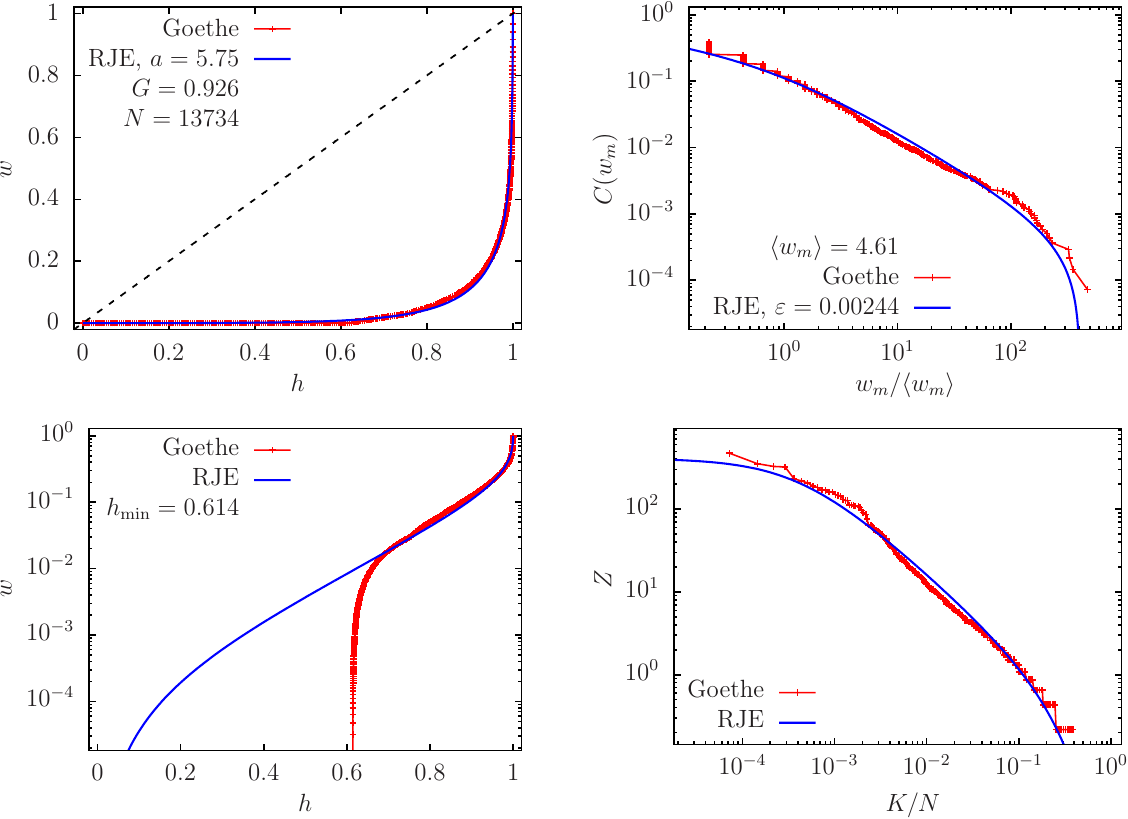}%
\end{center}
\caption{\label{figA7}
\label{fig_Goethe_multi4}
As Fig.~\ref{fig2} for Goethe.
}
\end{figure}

\begin{figure}[H]
\begin{center}
\includegraphics[width=0.9\columnwidth]{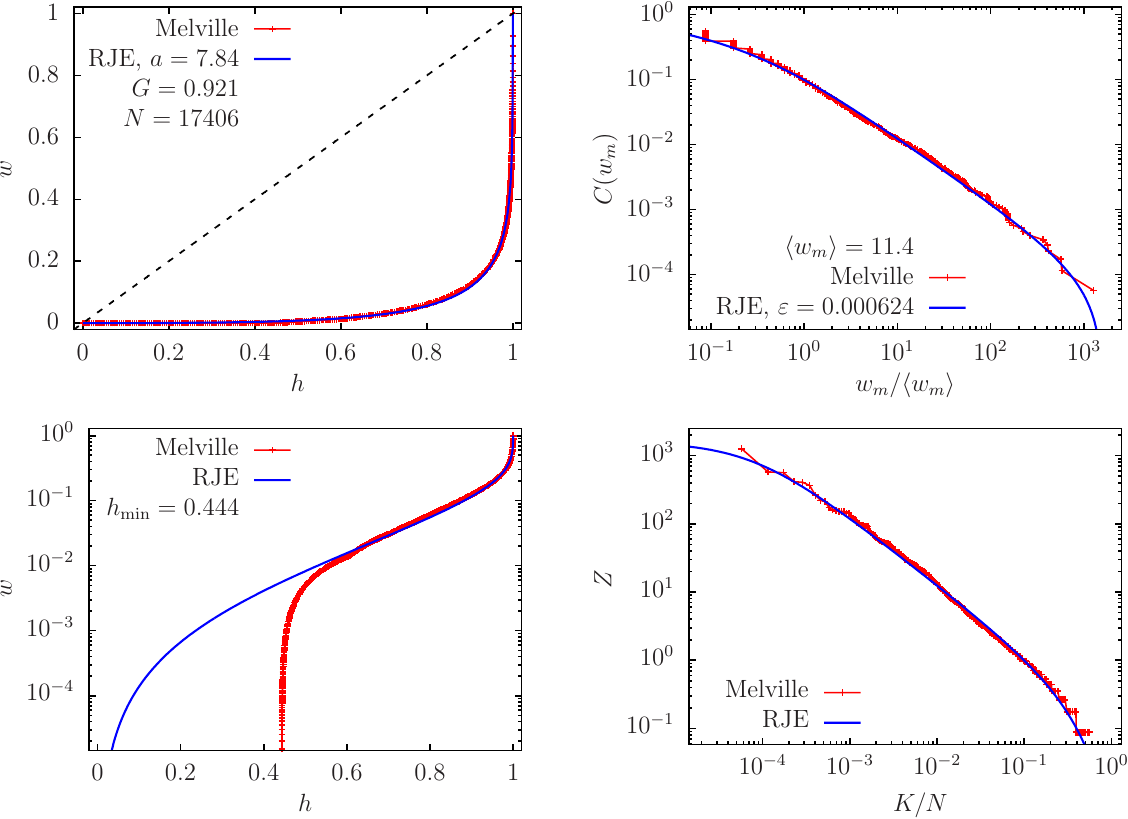}%
\end{center}
\caption{\label{figA8}
\label{fig_Melville_multi4}
As Fig.~\ref{fig2} for Melville.
}
\end{figure}

\begin{figure}[H]
\begin{center}
\includegraphics[width=0.9\columnwidth]{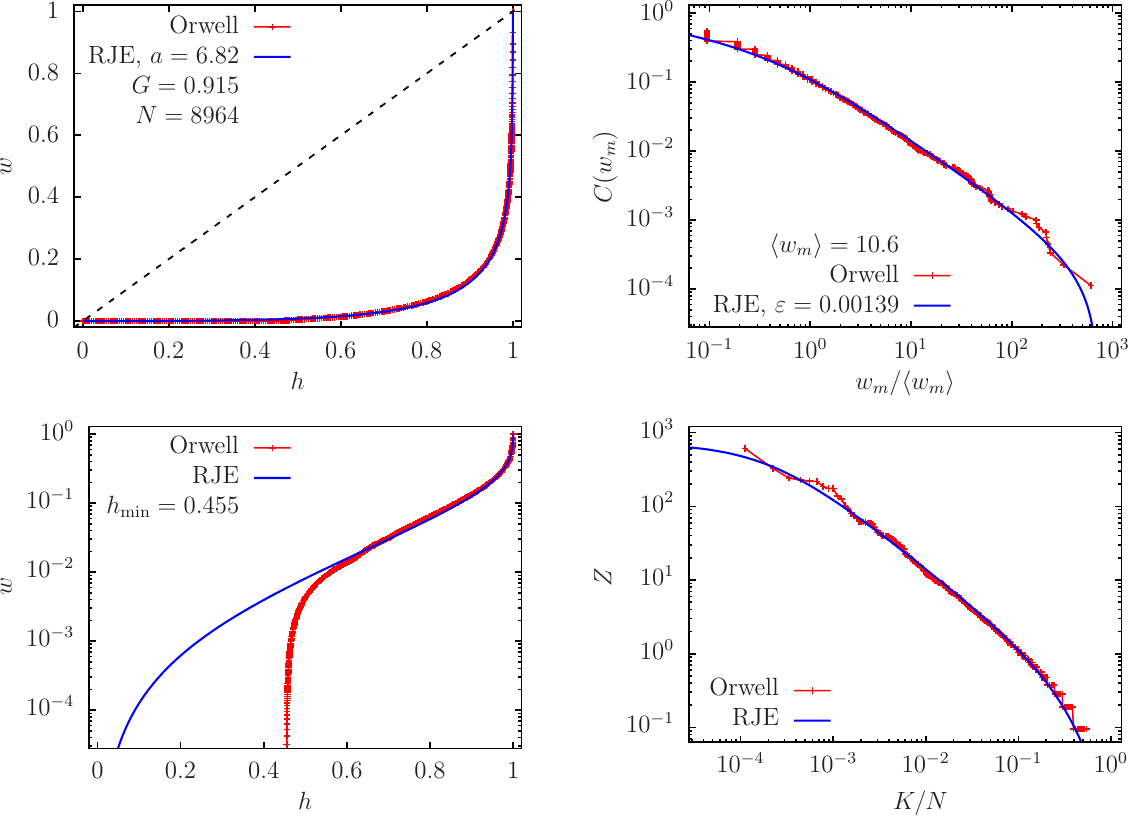}%
\end{center}
\caption{\label{figA9}
\label{fig_Orwell_multi4}
As Fig.~\ref{fig2} for Orwell.
}
\end{figure}

\begin{figure}[H]
\begin{center}
\includegraphics[width=0.9\columnwidth]{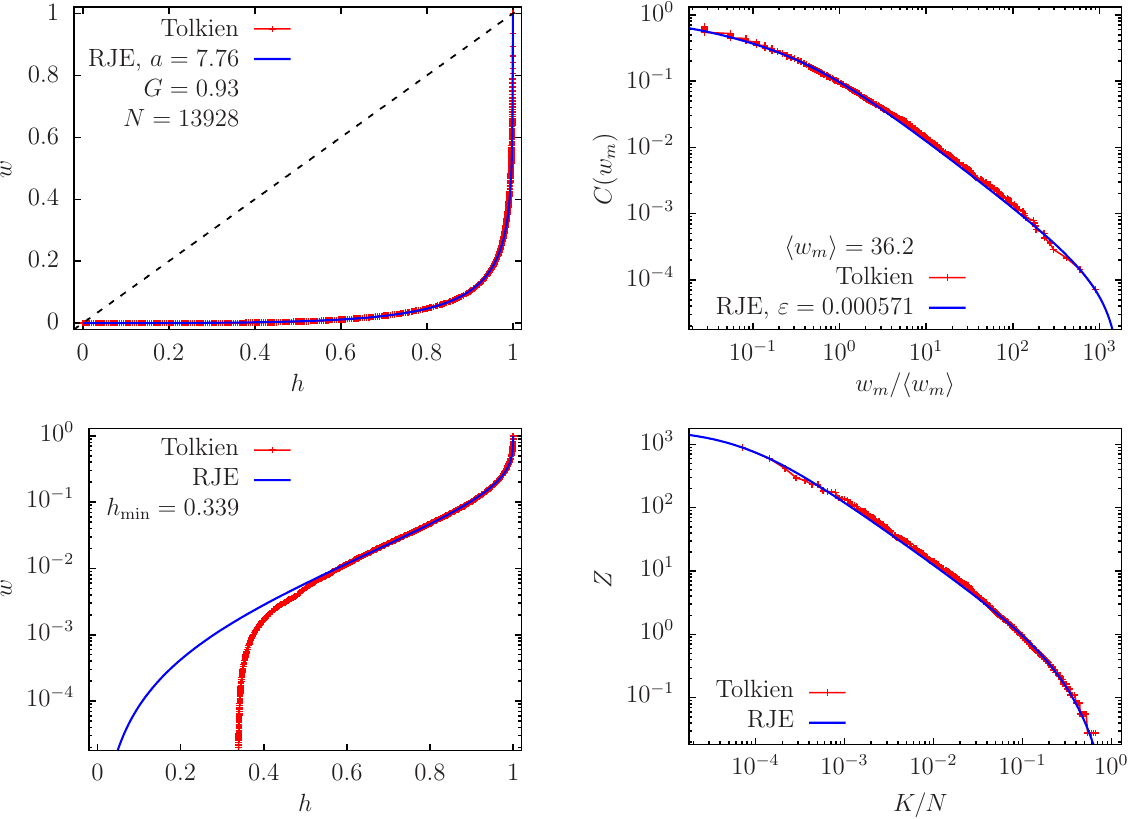}%
\end{center}
\caption{\label{figA10}
\label{fig_Tolkien_multi4}
As Fig.~\ref{fig2} for Tolkien.
}
\end{figure}

\begin{figure}[H]
\begin{center}
\includegraphics[width=0.9\columnwidth]{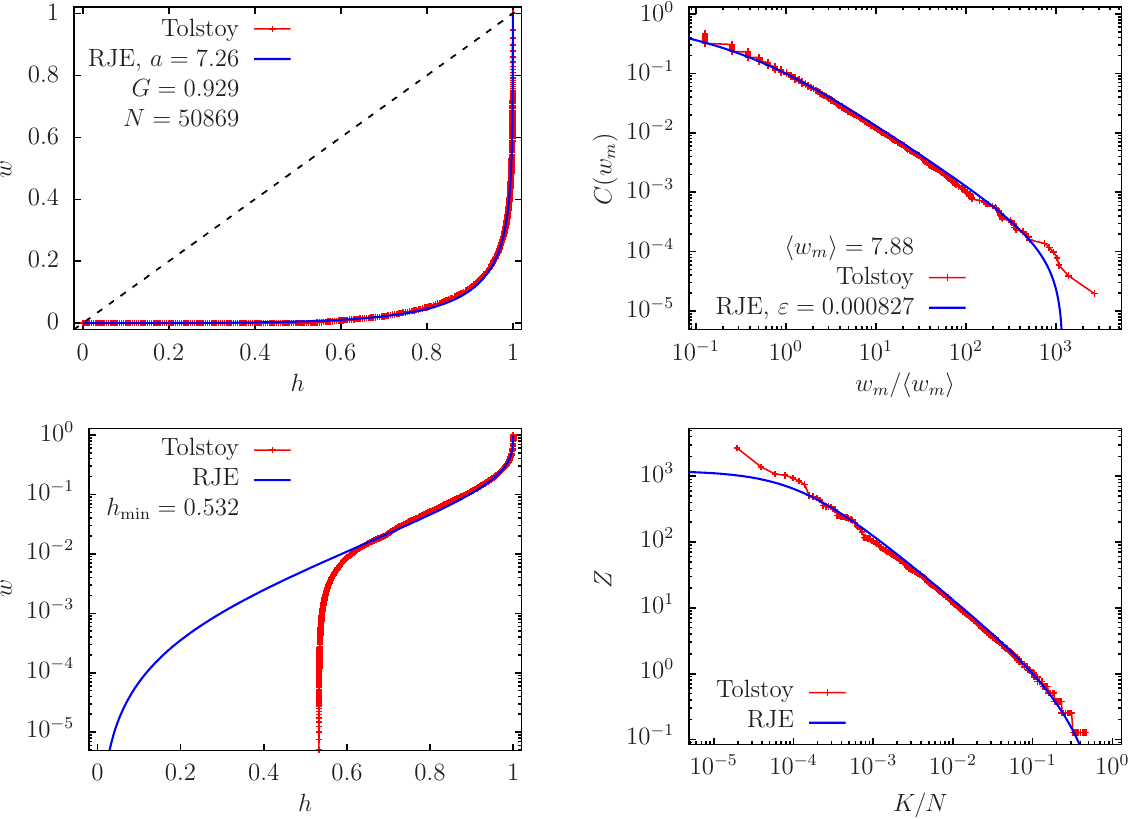}%
\end{center}
\caption{\label{figA11}
\label{fig_Tolstoy_multi4}
As Fig.~\ref{fig2} for Tolstoy.
}
\end{figure}

\begin{figure}[H]
\begin{center}
\includegraphics[width=0.9\columnwidth]{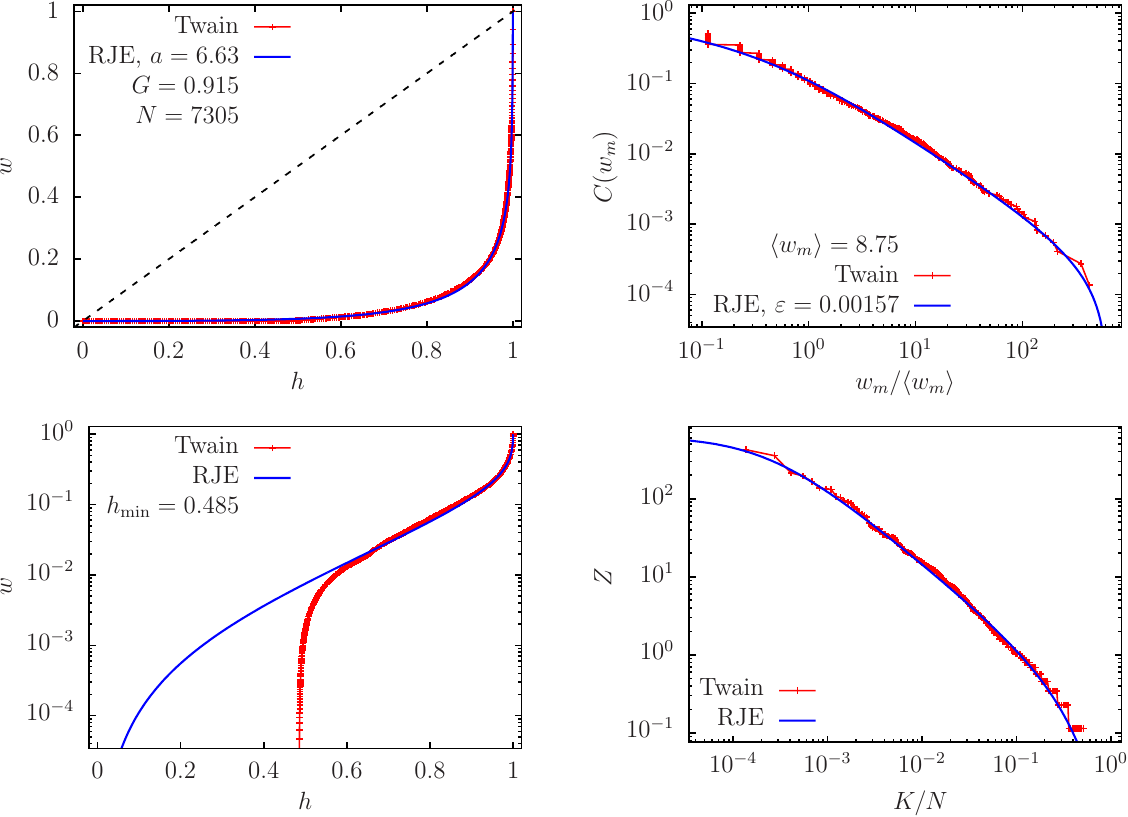}%
\end{center}
\caption{\label{figA12}
\label{fig_Twain_multi4}
As Fig.~\ref{fig2} for Twain.
}
\end{figure}

\begin{figure}[H]
\begin{center}
\includegraphics[width=0.9\columnwidth]{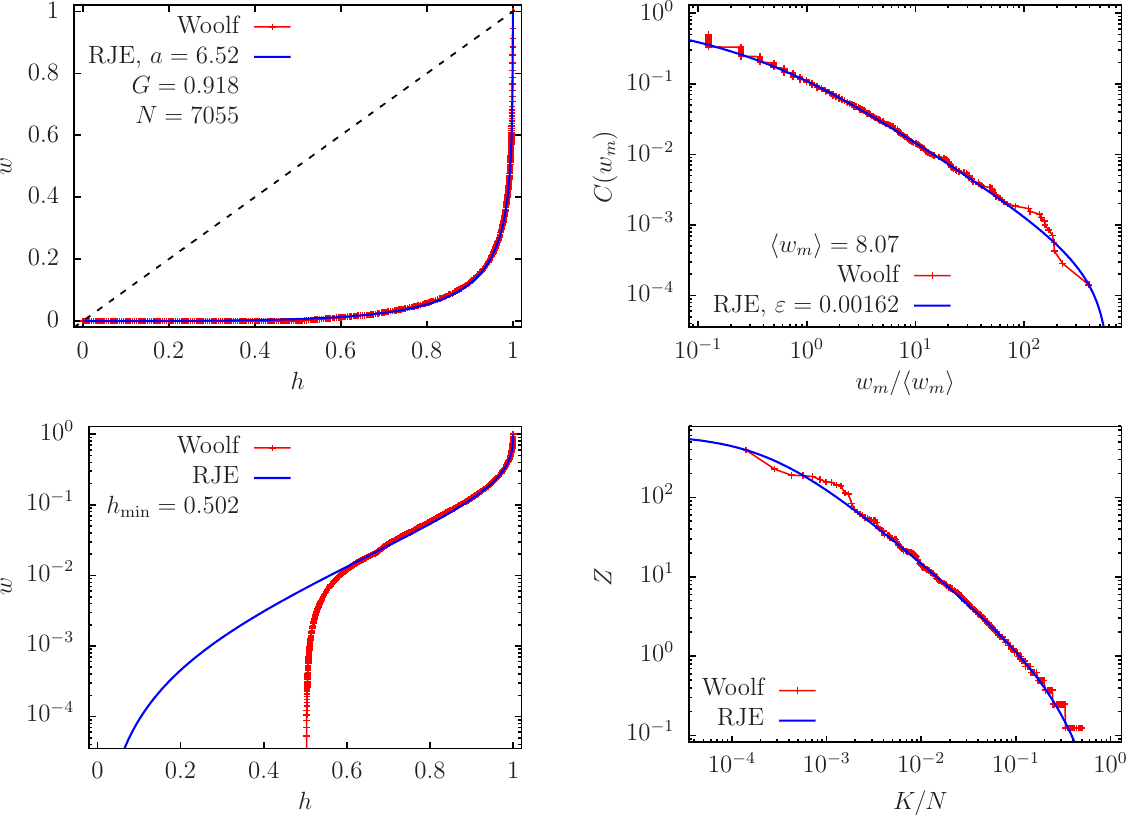}%
\end{center}
\caption{\label{figA13}
\label{fig_Woolf_multi4}
As Fig.~\ref{fig2} for Woolf.
}
\end{figure}

Here, we present Figs.~\ref{figA1}-\ref{figA13} which are in the 
same style as Fig.~\ref{fig2} 
for the cases of ANC, Asimov, Austen, Cervantes, Dante, Dickens, Goethe, 
Melville, Orwell, Tolkien, Tolstoy, Twain, and Woolf. 
These figures are mostly discussed in the main part of the text. 
Here, we point out that globally all real data curves (red data points) 
agree  very well with the corresponding RJE theory curves (blue lines) 
for optimal parameters $a$ and $\eps$ given in Figures for 
each data set. 

For the case of Tolstoy in Fig.~\ref{figA11}, 
there are some modest deviations at the 
tail of the Pareto curve at $C(w_m) < 10^{-4}$ 
(initial part of the Zipf curve) 
between data and the RJE theory. We note in analogy with the WTH concept
for the wealth inequality \cite{jsp,complex} that this tail
corresponds to the case of very rich households
(a lot of tokens for these words). Such households
or words of very high frequency are rare 
that leads to larger statistical fluctuation of such words
at the Pareto tail.

\subsection{Discussion of the limits of the Zipf law}

The analytical RJE approximation (\ref{eqCw2}) for the 
Pareto function, valid for $w_m\gg e^{-a}$ 
(assuming $|\mu|\ll e^{-a}$ which is well verified for all cases here), 
resembles partly a power law behavior except for the 
constant offset that ensures that $C_{\rm RJE}(1)=0$. 
If we assume in addition that also $w_m\ll 1$, we can neglect this offset 
which indeed gives the Zipf law, i.e. a decaying power law with exponent 
\footnote{See footnote 1 on the first page.}
$\alpha=1$:
\begin{align}
\label{eqCw3}
C_{\rm RJE}(w_m)\approx \left(\frac{\eps-\mu}{a}\right) w_m^{-1}
=A_{\rm RJE}\left(\frac{w_m}{\langle w_m\rangle}\right)^{-1}
\quad\mbox{if}\quad e^{-a}\ll w_m \ll 1
\ .
\end{align}
To have a significant interval of validity for this expression, we 
need rather larger values of 
$a$ (small values of $D^{-1}(a)\approx e^{-a}$). If this condition is verified 
the analytical Pareto curve of the RJE model is close to a decaying power 
law with exponent $\alpha=1$ for this interval. 

Concerning, the Zipf expression (\ref{Zeq1}), 
we can add the additional condition $K/N\gg A_{\rm RJE}\langle w_m\rangle 
\approx \eps/a$, such that we can neglect the constant term in the 
denominator which also gives the Zipf law
\begin{align}
\label{Zeq2}
Z&\approx A_{\rm RJE}\,\left(\frac{K}{N}\right)^{-1}
\quad\mbox{if}\quad \frac{\eps}{a}\ll \frac{K}{N}\ll \frac{\eps e^{a}}{a}\ .
\end{align}
This expression corresponds to the approximation (\ref{eqCw3}) for the 
Pareto function which are both valid for some intermediate interval 
in either $K$ or $w_m$. 

\begin{table}[h]
\caption{Table of power law fit parameters 
for the ANC data base and 15 books of writers. 
The columns correspond to name, the RJE power law coefficient 
$A_{\rm RJE}=(\eps-\mu)/(\eps a)\approx 1/a$, the Zipf fit parameters 
$A_{\rm Zipf}$, $\alpha$ 
and the min/max values of $Z=w_m/\langle w_m\rangle$ for the used 
fit interval. 
These parameters correspond either to the RJE power law behavior 
$C_{\rm RJE}(w)\approx A_{\rm RJE}(w_m/\langle w_m\rangle)^{-1}$ 
or the Zipf power law fit 
$K/N=C_{\rm data}(w_m)=A_{\rm Zipf}(w_m/\langle w_m\rangle)^{-1/\alpha}
=A_{\rm Zipf}Z^{-1/\alpha}$. 
The fit interval borders have been chosen 
as $Z_{\min}=2e^{-a}/\eps$ 
and $Z_{\max}=0.1/\eps$ and we expect 
the RJE power law behavior to be approximately valid 
for $Z_{\min}<Z<Z_{\max}$. The statistical fit errors of 
$A_{\rm Zipf}$ and $\alpha$ are well below 0.5\% for all cases. 
}
\label{tabA1}
\begin{center}
\begin{tabular}{lrrrrr}
\toprule
\textbf{Name} & $A_{\rm RJE}$ & $A_{\rm Zipf}$ & $\alpha$ & $Z_{\min}$ & $Z_{\max}$ \\
\midrule
ANC &  $0.0933$  &  $0.0847$  &  $1.07$  &  $1.73$  &  $3.93\times10^{3}$ \\
Asimov &  $0.172$  &  $0.129$  &  $1.09$  &  $2.06$  &  $35.1$ \\
Austen &  $0.129$  &  $0.101$  &  $1.12$  &  $1.69$  &  $198$ \\
Bible &  $0.111$  &  $0.0896$  &  $1.16$  &  $1.4$  &  $594$ \\
Cervantes &  $0.113$  &  $0.0917$  &  $1.09$  &  $1.45$  &  $530$ \\
Dante &  $0.155$  &  $0.113$  &  $1.08$  &  $2.23$  &  $70.6$ \\
Dickens &  $0.113$  &  $0.0894$  &  $1.17$  &  $1.2$  &  $417$ \\
Goethe &  $0.174$  &  $0.128$  &  $0.999$  &  $2.6$  &  $40.9$ \\
Hugo &  $0.153$  &  $0.103$  &  $1.1$  &  $2.54$  &  $88.5$ \\
Melville &  $0.128$  &  $0.0963$  &  $1.1$  &  $1.27$  &  $160$ \\
Orwell &  $0.147$  &  $0.115$  &  $1.06$  &  $1.57$  &  $71.9$ \\
Shakespeare &  $0.115$  &  $0.0877$  &  $1.14$  &  $1.4$  &  $441$ \\
Tolkien &  $0.129$  &  $0.107$  &  $1.11$  &  $1.49$  &  $175$ \\
Tolstoy &  $0.138$  &  $0.11$  &  $1.02$  &  $1.7$  &  $121$ \\
Twain &  $0.151$  &  $0.113$  &  $1.13$  &  $1.68$  &  $63.6$ \\
Woolf &  $0.154$  &  $0.123$  &  $1.07$  &  $1.82$  &  $61.7$ \\
\bottomrule
\end{tabular}

\end{center}
\end{table}

To test more explicitly the validity of these two expressions for real data, 
or more generally their power law properties, 
we perform for each data set a power law fit 
$C_{\rm data}(w)=A_{\rm Zipf}(w_m/\langle w_m\rangle)^{-1/\alpha}$ 
for the fit interval $Z=w_m/\langle w_m\rangle\in [2e^{-a}/\eps,0.1/\eps]$ 
where $a$ and $\eps$ are the optimal RJE parameters for the used data set and 
$\langle w_m\rangle$ is real data average of $w_m=f_m-1$ 
(see previous figures for specific values for each data set). 
Here, we choose the notation ``$-1/\alpha$'' for the fit 
exponent to be 
coherent with the notations of \cite{piantadosi} where 
$Z\sim w_m\sim f_m\sim K^{-\alpha}\sim C(w)^{-\alpha}$. 
We expect, that the analytical RJE power law approximations 
(\ref{eqCw3}) and (\ref{Zeq2}) are roughly valid for the proposed 
fit interval above. Therefore, it is reasonable to use the same interval 
(in the rescaled variable $Z=w_m/\langle w_m\rangle$) 
to perform a power law fit for the real data. (To be more precise 
the power law fit is actually done as the linear fit 
$y(x)=b+px$ with 
$y=\ln(C_{\rm data}(w))$, $x=\ln(w_m/\langle w_m\rangle)$, 
$b=\ln(A_{\rm Zipf})$ and 
$p=-1/\alpha$.)

In Table~\ref{tabA1}, we provide for each data set the 
RJE power law coefficient $A_{\rm RJE}$ defined in Eq. (\ref{ARJEdef}), 
the two fit parameters 
$A_{\rm Zipf}$, $\alpha$ and also the numerical values of 
the borders of the fit interval (in units of $\langle w_m\rangle$, 
i.e. for $Z=w_m/\langle w_m\rangle$). 
The obtained values of $\alpha$ are typically in the interval 
$[1,1.17]$ with most values $\alpha\approx 1.1$ and minimal value 
$\alpha=0.999$ for Goethe. 
The coefficients $A_{\rm RJE}\approx 1/a$ and $A_{\rm Zipf}$ 
are quite close with roughly comparable values. 

\begin{figure}[h]
\begin{center}
\includegraphics[width=0.9\columnwidth]{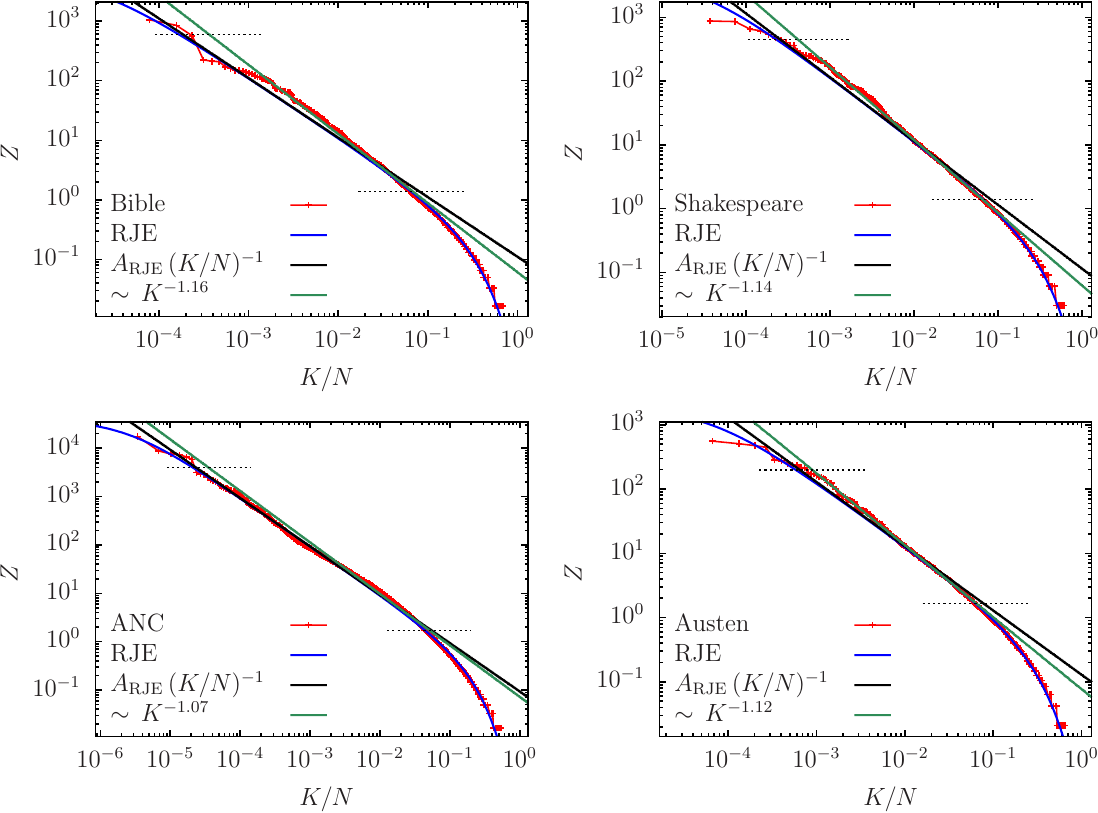}%
\end{center}
\caption{\label{figA14}
\label{fig_Fit4}
Graphical illustration of the power law fits according to 
Table~\ref{tabA1} for the cases of Bible, Shakespeare, ANC and Austen. 
Shown are the Zipf curves of $Z$ versus the rescaled rank $K/N$ 
for the word data (red data points) and the precise optimal 
RJE curve (blue line). 
The two straight lines correspond to the analytical RJE power law 
approximation $Z=A_{\rm RJE}(K/N)^{-1}$ 
(black line) and the power law expression $Z=(K/(N A_{\rm Zipf}))^{-\alpha}$ 
(green line) obtained from the fit $K/N=A_{\rm Zipf} Z^{-1/\alpha}$. 
The thin horizontal black dotted lines indicate 
the values of $Z_{\min}$ and $Z_{\max}$ 
for the used fit interval given in Table~\ref{tabA1}. }
\end{figure}

In Fig.~\ref{figA14}, we provide a graphical illustration of this 
power law fit for the Zipf curves of the data for the four examples of 
Bible, Shakespeare, ANC and Austen. As in the previous figures, 
the real word data corresponds to 
the red data points and the exact RJE curve (for optimal values 
$a$ and $\eps$ for the given data set) to the blue line. 
The analytical RJE power law approximation 
(\ref{Zeq2} with exponent ``$-1$'' is shown as the black line and the 
above power law fit corresponds to the green line. In the figure panels, 
also the values of $Z_{\min}=2e^{-a}/\eps$ and $Z_{\max}=0.1/\eps$ 
for the used fit interval (in the Zipf variable $Z$) are marked as 
thin dotted black lines.

First, we observe that the blue line (exact RJE expression 
obtained as the inverse function of (\ref{eqCw1})) 
and black line (analytic RJE power law approximation (\ref{Zeq2})) are quite 
close for the values in the fit interval which is expected. There are 
slight deviations between them at the interval boundaries which is due to 
the fact that the chosen interval is slightly on the ``larger'' side 
and that the conditions of validity of (\ref{Zeq2}) are not perfectly 
verified at its boundaries. The green line for the 
power law fit is rather close to the black line but there are some visible 
deviations and the slope of the green line corresponds to values of 
``$-\alpha$'' where $\alpha$ is slightly larger than unity. The green fit 
line agrees typically very well with most of the (red) data points 
inside the fit interval except for a small number of data points at the 
top of the interval where the density of data points is also reduced 
and which have a smaller influence on the fit result. As already observed 
in the previous figures, the blue RJE curve agrees globally quite well with 
the data nearly for all data points, also outside the fit interval and 
in particular in the range of large $K/N>0.1$ 
where both data and precise RJE theory are very different 
from the power law behavior. Concerning the range of smallest values 
of $K\le 5$-$10$ (top frequent words with largest values of $w_m$), 
the data points are not very far from the black line for the cases 
of Bible and ANC while for Shakespeare and and Austen they 
deviate a bit more from both blue and black RJE curves (and even more from 
the green fit line). This region is statistically not very reliable and 
concerns only a few data points. However, also here the blue RJE curve still 
provides the best agreement in comparison to the other two power law 
lines. 

We mention, that the more refined approximation (\ref{Zeq1}) (not shown 
in the figure), which is valid for $K/N\ll \eps e^{a}/a$, agrees quite 
precisely with the exact blue RJE curve for smaller values of $K/N<10^{-2}$ 
and with the black line for larger values of $K/N>10^{-2}$. 

We have also tested the case of a reduced fit interval, 
i.e. $Z_{\min}=4e^{a}/\eps$ and $Z_{\max}=0.05/\eps$ (factor 2 modification 
of both interval borders to the inside). In this case, 
the obtained fit values of $A_{\rm Zipf}$ and $\alpha$ are still quite 
similar. However, the values of $\alpha$ 
are generally a bit closer to unity, e.g. for the four cases of 
Fig.~\ref{figA14}, we have $\alpha=1.12,1.15, 0.993,1.1$ instead 
of the values $\alpha=1.16,1.14,1.07,1.12$ in Fig.~\ref{figA14} for 
the initial fit interval. For the reduced interval, the black and green lines 
are typically a bit closer. 

In summary, we see that the RJE Zipf curves exhibit 
for the cases of rather large values of $a$ and small values of $\eps$ 
the Zipf power law in some intermediate interval, 
$\eps/a\ll K/N\ll \eps e^{a}/a$, which can be quite large. 
However, for the regions outside this interval there are significant 
deviations from this power law, and as we can see in 
Fig.~\ref{figA14} and also in the previous figures, 
the RJE theory curves fit remarkably well 
the real word distribution data for most cases and the full range 
with only two free 
parameters being $\eps$ and $a$. We note, that typical power law type fits 
of \cite{piantadosi} are done with at least two parameters being 
the exponent and the numerical coefficient. In certain cases, a third parameter 
$\beta$ for a shifted power law, $f_m\sim Z\sim (\beta+K)^{-\alpha}$, 
is added which is similar to the offset term 
in either (\ref{eqCw2}) or (\ref{Zeq1}) but there this term is 
simply fixed by the theory and does not add a third fit parameter. 
We note that even with this extension, there are still significant 
deviations with the data (used here in this work) 
for larger values of $K/N>0.1$.

The above discussion focuses on the Zipf curves but the situation is very 
similar for Pareto curves where the roles of $x$- and $y$-axis are 
exchanged. However, 
it is quite complicated to construct potential Lorenz curves from simple 
power law Pareto functions due to their singular behavior for small 
and large values of $w_m$ and the necessary integral to compute the 
cumulated wealth variable. 
For the more refined shifted power law approximation (\ref{eqCw2}) 
for the Pareto function, it is still possible to reconstruct the Lorenz curve, 
using a simple procedure given in \cite{complex}, which gives 
the formula $w(h)\approx \ln\left(1-h/(1+C_2)\right)/
\ln\left(C_2/(1+C_2)\right)$ with $C_2=(\eps-\mu)/a$. However, this expression 
does not agree well with neither the real data nor 
the precise RJE Lorenz curve (\ref{eqwofh}) except for the 
very rich phase at $h\approx 1$. On the other hand, 
the precise RJE Lorenz curve 
typically fits the data Lorenz curve extremely well as can be seen in the 
above figures. 
The reason for this is that the Lorenz curve emphasizes the poor phase 
(large values of $K$) where the deviations between power law and 
data are strongest while the RJE theory agrees quite well with the data 
in this region.

\starttwocolumn

\begin{acknowledgments}
The authors acknowledge support from 
NANOX $N^\circ$ ANR-17-EURE-0009 in the framework of 
the Programme Investissements d'Avenir (project MTDINA).
\end{acknowledgments}

\bibliographystyle{compling}
\bibliography{references}

\end{document}